\documentclass[journal]{IEEEtran}

\ifCLASSINFOpdf
\else
\fi
\usepackage{graphicx} 
\usepackage{svg}
\usepackage{amsmath,amsfonts}
\usepackage{amssymb}
\usepackage{algorithmic}
\usepackage{algorithm}

\usepackage{booktabs}
\usepackage{cite}
\usepackage{bm}
\usepackage{float}
\usepackage{multirow}
\usepackage{color}
\usepackage{caption}
\usepackage{epstopdf}
\usepackage{fixltx2e}
\usepackage{longtable}
\usepackage{diagbox}
\usepackage{changepage}
\usepackage{comment}
\usepackage{cases}
\usepackage{makecell}
\usepackage{mathabx}

\usepackage{array,threeparttable}

\begin{document}
%
\title{Efficient Twin Migration in Vehicular Metaverses: Multi-Agent Split Deep Reinforcement Learning with Spatio-Temporal Trajectory Generation}

\author{Junlong Chen, Jiawen Kang, Minrui Xu, Fan Wu, Hongliang Zhang,  Huawei Huang, \\ Dusit Niyato, \IEEEmembership{Fellow,~IEEE}, Shiwen Mao, \IEEEmembership{Fellow,~IEEE} 
\thanks{Junlong Chen, Jiawen Kang, and Shengli Xie are with the School of Automation, Guangdong University of Technology
, Guangzhou, China (e-mail: 3121001036@mail2.gdut.edu.cn; kjwx886@163.com; Shlxie@gdut.edu.cn).}
\thanks{Minrui Xu and Dusit Niyato are with the School of Computer Science and Engineering, Nanyang Technological University, Singapore, Singapore (e-mail: minrui001@e.ntu.edu.sg; DNIYATO@ntu.edu.sg).}
\thanks{Fan Wu is with the School of Computer Science and Engineering, Central South University, Changsha, China (e-mail: wfwufan@csu.edu.cn).}
\thanks{Hongliang Zhang is with the School of Electronics, Peking University, Beijing, China (e-mail: hongliang.zhang92@gmail.com).}
\thanks{Huawei Huang is with the School of Data and Computer Science, Sun Yat-Sen University, Guangzhou, China (e-mail: huanghw28@mail.sysu.edu.cn).}
\thanks{Shiwen Mao is with the Department of Electrical and Computer Engineering, Auburn University, Auburn, USA (e-mail: smao@ieee.org).}

}

\markboth{Journal of \LaTeX\ Class Files,~Vol.~14, No.~8, August~2015}%
{Chen \MakeLowercase{\textit{et al.}}: Bare Demo of IEEEtran.cls for IEEE Journals}

\maketitle

\begin{abstract}
Vehicle Twins (VTs) as digital representations of vehicles can provide users with immersive experiences in vehicular metaverse applications, e.g., Augmented Reality (AR) navigation and embodied intelligence. VT migration is an effective way that migrates the VT when the locations of physical entities keep changing to maintain seamless immersive VT services. However, an efficient VT migration is challenging due to the rapid movement of vehicles, dynamic workloads of Roadside Units (RSUs), and heterogeneous resources of the RSUs. To achieve efficient migration decisions and a minimum latency for the VT migration, we propose a multi-agent split Deep Reinforcement Learning (DRL) framework combined with spatio-temporal trajectory generation. In this framework, multiple split DRL agents utilize split architecture to efficiently determine VT migration decisions. Furthermore, we propose a spatio-temporal trajectory generation algorithm based on trajectory datasets and road network data to simulate vehicle trajectories, enhancing the generalization of the proposed scheme for managing VT migration in dynamic network environments. Finally, experimental results demonstrate that the proposed scheme not only enhances the Quality of Experience (QoE) by 29\% but also reduces the computational parameter count by approximately 25\% while maintaining similar performances, enhancing users' immersive experiences in vehicular metaverses.

\end{abstract}


\IEEEpeerreviewmaketitle

\section{Introduction}\label{introduction}












Vehicular metaverses, as a novel concept in advanced transportation systems, are designed to offer engaging virtual-physical interactions for both drivers and passengers within a 3D digital environment using smart vehicles. In vehicular metaverses, Vehicle Twins (VTs), as digital representations of physical vehicles, emerge as pivotal components to craft immersive experiences through applications ranging from AR navigation aids for tourists to embodied intelligence, engaging 3D entertainment games~\cite{luo2023privacy}. Specifically, VT services' seamless integration and interaction for maintaining vehicular metaverses depends on unparalleled rendering precision and minimal latency to ensure the immersive user experience~\cite{li2023digital}. Nevertheless, the inherent rapid movement of vehicles and the diverse computing resources and workload of RSUs present significant challenges to maintaining immersive experiences provided by VT services~\cite{chen2024aiot}.

The landscape of vehicular metaverses is fraught with complexities. Notably, the rapid movement of vehicles brings significant challenges in maintaining dynamic and seamless VT services~\cite{zhong2023blockchain}. VT migration is an effective way that migrates the VT when the locations of physical entities keep changing to maintain seamless immersive VT services. Nevertheless, traditional methods for VT migration struggle to keep pace with the rapid changes in user location, resulting in latency issues that detract from the immersive experience~\cite{hui2023digital}. Furthermore, the dynamic workload of RSUs and their diversity deployment exacerbate the difficulty in ensuring efficient VT migrations. These challenges emphasize the need for a more sophisticated solution that can adapt to the fast-moving and varying environments of vehicular networks, ensuring immersive VT services.

To tackle the VT migration decision problem in vehicular metaverses, Multi-Agent Deep Reinforcement Learning (MADRL), an extension of DRL that caters to environments involving multiple agents, emerges as a powerful solution~\cite{chen2024aiot}. MADRL enables decentralized agents, e.g., vehicles and RSUs, to learn adaptive migration strategies under dynamic conditions, such as changing vehicle positions and fluctuating resource availability. Although MADRL is effective in handling these complex decision-making tasks, it also introduces challenges related to heavy computation demands, privacy, and efficient convergence.


To address these issues, we integrate split learning~\cite{otoum2022feasibility} into the MADRL framework. Split learning achieves this by partitioning a deep neural network across multiple nodes, e.g., processing initial layers locally at vehicles and the subsequent layers at RSUs. Split learning excels in processing sensitive data locally at vehicles while offloading heavier computations to RSUs, optimizing bandwidth by transmitting only intermediate results rather than raw data. This method efficiently manages the computational load and preserves data privacy by processing sensitive information close to its source.


The synergy of split learning allows for dynamic adaptation to changing network conditions and resource availability, optimizing system performance without compromising privacy or responsiveness. By coupling the local computation capabilities of split learning with the strategic coordination provided by MADRL, our scheme effectively tackles the complexities of VT migration, leading to improved efficiency and user experience in vehicular metaverses.

In addition, trajectory generation leverages historical vehicle data to generate new trajectories that reflect the spatial and temporal mobility of vehicles, addressing the challenges of scarcity and high costs associated with collecting long-distance trajectory data over various traffic~\cite{chen2021trajvae,luo2022estnet}. However, effectively generating realistic traffic trajectories presents challenges, including the need to capture both spatial and temporal dependencies accurately. It is crucial to ensure that the generated trajectories not only mimic real-world vehicle movements but also exhibit plausible traffic flow patterns across different regions and time periods. By simulating realistic traffic patterns, the trajectory generation provides enriched environments~\cite{wang2023trajectory} that aid MADRL in modeling complex interactions among vehicles and RSUs that are crucial for VT migrations within vehicular metaverses.

The integration of trajectory generation with MADRL allows our framework to simulate intricate vehicular dynamics effectively. This facilitates the learning of DRL agents, enhancing VT migration strategies that are both adaptable and efficient across diverse real-world conditions. Consequently, our scheme not only mitigates the issue of data scarcity but also significantly boosts the efficiency and reliability of VT migrations, optimizing user experiences in vehicular metaverses.

The contributions of this paper are outlined as follows.
\begin{itemize}
\item To provide immersive VT services to vehicular users, we proposed an innovative framework that integrates VT service latency and quality of service to maximize the Quality of Experience (QoE) of vehicular users in vehicular metaverses. By formulating the problem as a Partially Observable Markov Decision Process (POMDP), we utilize MADRL to dynamically provide migration decisions to minimize VT service latency while maximizing the VT service quality, thus maximizing the QoE of vehicular users.

\item To tackle the challenges arising from the rapid movement of vehicles, the animated changes in network topology, and the spatio-temporal variability in RSU deployment and workload, we propose an Efficient Spatio-Temporal Trajectory Generation (EST-TG) algorithm that generates realistic vehicular traffic scenarios, reducing the dependence on extensive real-life data collection. Experimental results show that the proposed EST-TG algorithm not only facilitates vehicle moving pattern simulation but also enhances the generalization and adaptability of the MADRL algorithm across varying traffic conditions.

\item With the generated data, we propose a novel Multi-agent Split Reinforcement Learning (MSRL) algorithm that dynamically switches decision networks to accommodate the inconsistent deployment of RSUs and the demands of high-resource VT tasks. By integrating split learning and MADRL, the proposed scheme leverages both the local computational capabilities and edge computational capabilities to enhance the decision efficiency of VT pre-migration. Numerical results show that our scheme outperforms other baselines in maximizing the QoE of vehicular users while reducing computational parameters. 

\end{itemize}

The organization of this paper includes the following sections: Section \ref{related} reviews the related work. Section \ref{model} outlines the system model and problem formulation. The proposed MSRL scheme for pre-migration decisions is detailed in Section \ref{madrl}. In Section \ref{Gen}, we explain the EST-TG algorithm for VT pre-migration decisions. The numerical results are presented in Section \ref{result}, and the paper concludes with Section \ref{conclusion}.

\section{Related Work}\label{related}

\subsection{Immersive Service in Vehicular Metaverses}

The evolution of immersive services in vehicular metaverses has been significantly influenced by advancements in digital twins and AI technologies. These technologies initially facilitated the creation of synchronized virtual representations of physical entities, directly enhancing user experiences by enabling real-time interaction between the physical and virtual realms \cite{han2022dynamic}. Subsequent works have integrated edge computing to further improve this synchronization, which is crucial for deepening the immersion of vehicular metaverse services \cite{xu2022full}. Recent works have explored the use of generative AI to simulate realistic autonomous driving scenarios, enhancing the alignment between physical vehicles and their digital counterparts, thus enriching the overall immersive experience \cite{xu2023generative}. Additionally, a learning-based framework proposed by Chen \textit{et al.} addresses service migration within these metaverses, focusing on managing the dynamic and resource-constrained environments typical of vehicular networks to optimize user experiences \cite{chen2023multiple}. However, there remains a gap in addressing the dynamic decision-making challenges posed by limited decision resources in vehicular metaverses.

\subsection{Trajectory Generation}
Trajectory generation in vehicular networks has evolved from basic route mapping techniques to sophisticated algorithms incorporating Generative Adversarial Networks (GANs). Raja \textit{et al.} \cite{raja2022intelligent} established foundational methodologies for generating trajectories in 6G networks, focusing on efficiency and precision in path planning. Building on this, Zheng and Mangharam \cite{zheng2023differentiable} developed differentiable trajectory generation for car-like robots, employing Radial Basis Function Networks to enhance adaptability and accuracy in vehicular movement predictions. Li and Zhao \cite{li2023trajectory} addressed ultra-low-frequency travel routes within intricate road networks, improving solutions for large-scale trajectory planning. Jiang \textit{et al.} \cite{wang2021large} introduced two-stage GANs, a breakthrough for generating large-scale trajectories and simulating dynamic vehicular motions more realistically. However, these methods often only focus on temporal data and require intensive computing resources to train the model, lack the capacity to adapt to real-time changes in dynamic vehicular environments. Our scheme overcomes these shortcomings by leveraging real-time data and efficient analytical models, significantly enhancing the support for immersive services in vehicular metaverses by accurately simulating dynamic vehicular movements.

\subsection{Reinforcement Learning in Service Migration}
The integration of reinforcement learning (RL) into service migration strategies within mobile and edge computing environments has emerged as a key area of innovation. By leveraging RL, these strategies aim to optimize the allocation of computational resources and minimize latency through dynamic adaptation to shifting network conditions and fluctuating user demand. For instance, Afrasiabi \textit{et al.} \cite{afrasiabi2022reinforcement} developed an RL-based optimization framework for application component migration in Cloud-Fog settings, illustrating how RL can continuously adjust resource allocation to maintain service continuity and improve overall resource utilization. Similarly, Dong \textit{et al.} \cite{dong2023dynamic} proposed a dynamic RL-driven service migration approach for mobile edge computing, demonstrating the ability of RL to make predictive decisions that proactively bring services closer to end-users, thereby reducing latency and enhancing user experience amid changing network dynamics.

More recently, multi-agent deep reinforcement learning (MADRL) techniques have been introduced to further improve service migration decisions by enabling distributed, cooperative learning among multiple agents. For example, Chen \textit{et al.} \cite{chen2024aiot} presented a MADRL-based service migration scheme that adapts to the rapidly evolving conditions of vehicular metaverses, utilizing multiple agents to collectively maintain continuous and immersive services even as vehicles move and network conditions shift. While this approach shows promise in handling complex, dynamic environments, it does not fully address the increasing computational demands associated with real-time decision-making in such settings.

Building on these insights, our proposed solution integrates split learning with a MADRL algorithm. This integration not only decreases decision-making latency but also enhances the security of the user data that is used in decision-making, providing a seamless and improved user experience.

\section{System Model and Problem Formulation}\label{model}
In this section, we present the VT migration framework in the vehicular metaverse. First, we introduce models for communication, migration, and computing. Subsequently, we expound upon the total latency associated with VT task migration.

\begin{figure}[t]  	
\centerline{\includegraphics[width=0.5\textwidth]{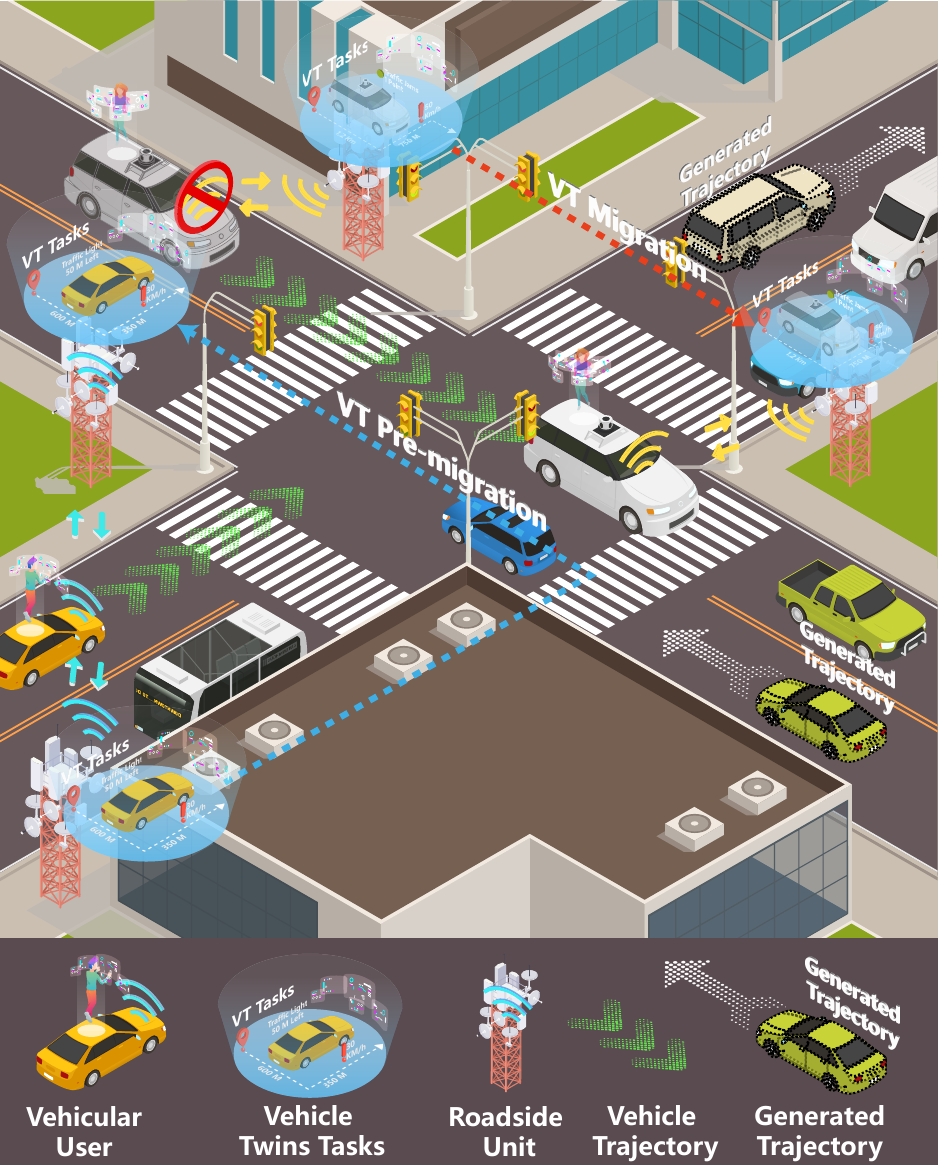}}  	
\caption{The migration of VTs tasks with generated trajectories in the vehicular metaverse.} \label{fig: system}  
\end{figure}

The proposed VT migration system model is illustrated in Fig.~\ref{fig: system}. The system comprises multiple RSUs and vehicular users, with the RSU set denoted as $\mathcal{E}=\{1, ..., e, ..., E\}$ and the vehicular user set denoted as $\mathcal{V}=\{1, ..., v, ..., V\}$. Each RSU, represented by $e$, has GPU computing resources $C_e$, maximum workload $L_e^{max}$, uplink bandwidth $B^{u}_e$, and downlink bandwidth $B^{d}_e$ for transmitting input data of VT tasks, receiving data requests from vehicular users, and transmitting VT task results. In this system, time is divided into time slots, denoted as $T=\{1, ..., t, ..., T_{max}\}$. At time slot $t$, vehicular users can transmit VT task requests to the currently serviced RSU, for instance, information required for updating the state of VT in the virtual space. Additionally, vehicular users can proactively migrate a portion of VT tasks to specific available RSUs by executing action $a_v(t) \in a$ for advanced processing. To enhance the rationality of the pre-migration decision and to improve the QoE of the service, each vehicular user can make the decision based on the information of its current location, the load of the nearby RSUs, and the number of requests.

\subsection{Communication Model}
The process of determining communication latency begins when VT tasks are transmitted through wireless communications between vehicles and RSUs. The location of RSU $e$ is denoted as $P_e = (x_e,y_e)$. Meanwhile, the coordinates of vehicular user $v$ change over time and are represented as $P_v(t) = [x_v(t), y_v(t)]$ during the time slot $t$. Consequently, the Euclidean distance separating RSU $e$ and vehicular user $v$ at time slot $t$ is given by $d_{v,e}(t) = \sqrt{|x_e - x_v(t)|^2 + |y_e - y_v(t)|^2}$, with $|\cdot|$ denoting the Euclidean distance function \cite{Lee1993Euclid}.

In wireless communications, the latency caused by offloading a VT task request to RSU $e$ during time slot $t$ is influenced by the uplink transmission rate, which can be calculated as~\cite{shannon2001mathematical} $R^{u}_{v,e}(t)=B^{u}_e\log_2[1+\frac{p_v h_{v,e}(t)}{\sigma^2_e}]
$, where $p_v$ represents the transmit power of vehicular user $v$, $\sigma_e$ denotes the additive Gaussian white noise at RSU $e$, and indicates the wireless uplink and downlink channels between vehicular user $v$ and RSU $e$. In vehicular networks, the Rayleigh fading channel $h_{v,e}(t) = A(\frac{c}{4\pi f d_{v,e}(t)})^2$ is considered, where $A$ denotes the channel gain coefficient, $f$ stands for the carrier frequency, $d_{v,e}(t)$ denotes the Euclidean distance separating vehicular user $v$ from RSU $e$ during time slot $t$, and $c$ is the speed of light. To gain access to immersive VT services, vehicle user $v$ is required to send VT task requests to the active RSU. The request size $s_v^{r}(t)$ is determined by user demands. Consequently, at time slot $t$, the latency of transmitting a VT task request to the currently serviced RSU can be calculated as \begin{equation}T^{u}_{v,e}(t)=\frac{D_v^{u}(t)}{R^{u}_{v,e}(t)}.\label{eq2}
\end{equation}

Since a VT task can be processed on multiple RSUs, vehicular users need to receive the processing results of the same VT task from various RSUs. Specifically, when a VT task request reaches an RSU, the RSU creates a VT task with a size $D^{task}_v(t)$ according to the VT task request. For vehicular users, they can choose to pre-migrate part of the VT task with a size $D^{m}_v(t)$ to the specified RSU $e_{m}$. Furthermore, after RSUs have completed processing the VT tasks, the results are transmitted back from multiple RSUs to the vehicular users through vehicular networks. Analogous to the uplink transmission rate, the rate of downlink transmission for sending VT task results from RSUs to vehicular user $v$ can be estimated as $R^{d}_{v,e}(t)=B_e^{d}\log_2(1+\frac{p_v h_{v,e}(t)}{\sigma^2_e}).$ Therefore, the transmission latency of transmitting VT task results from RSUs to vehicular user $v$ at time slot $t$ is defined as 
\begin{equation}
T^{d}_{v}(t)=\sum_{e=1}^{E} \frac{D^{d}_{v,e}(t)}{R^{d}_{v,e}(t)},\label{eq3}
\end{equation}
where $D^{d}_{v,e}(t)$ is the size of the VT task results transmitted from RSU $e$.

\begin{table}[t]

	\begin{threeparttable}
        \setlength{\abovecaptionskip}{0pt}
	\caption{Primary Symbols Utilized in This Paper}
	\begin{tabular}{c || p{6.5cm}}
		\hline
		\textbf{Notation} & \textbf{Description} \\
		\hline
  
        $B^{u}_e$ & The uplink bandwidth of RSU $e$\\
        $B^{d}_e$ & The downlink bandwidth of RSU $e$\\
        $B_{e,e_{m}}$ & The physical bandwidth of between current serving RSU $e$ and pre-migrated
RSU $e_m$\\

        $C_e$ &  The GPU computing capacity available at RSU $e$  \\
        $C_{e_{m}}$ & The GPU computing capacity available at RSU $e_{m}$\\

        $L_e(t)$ &  The current workload of RSU $e$  \\
		$L_e^{max}$ & The maximum workload of RSU $e$ \\

        $R^{u}_{v,e}(t)$ &  The uplink transmission rate between vehicular user $v$ and RSU $e$ \\
		$R^{d}_{v,e}(t)$ &  The downlink rate transmission rate between vehicular user $v$ and RSU $e$ \\

		$D^{task}_{v}(t)$ & The total VT task size of vehicular user $v$\\ 
		$D^{m}_{v}(t)$ & The portion of the VT task that is pre-migrated of vehicular user $v$\\
		$D^{L}_{v}(t)$ & The remaining VT task size at RSU $e$\\

		$T^{u}_{v,e}(t)$ & The latency of transmitting a VT task request of vehicular user $v$ to RSU $e$\\
		$T^{d}_{v}(t)$ & The latency of transmitting VT task results of vehicular user $v$\\
		$T^{m}_{v}(t)$ & The latency of pre-migrating VT task of vehicular user $v$ between RSUs\\
	
		$T^{p}_{v,e}(t)$ & The processing latency of VT task of vehicular user $v$ at RSU $e$\\
        $T^{p}_{v,e_{m}}(t)$ & The processing latency of VT task of vehicular user $v$ at pre-migrated RSU $e_{m}$\\
        $T^{p}_{v}(t)$ & The parallel processing latency of VT task of vehicular user $v$\\
		$T^{total}_v(t)$ & Total latency of VT services of vehicular user $v$\\
        $QoE_v(t)$ & The quality of VT service experience of vehicular user $v$\\
        $a_v(t)$ & The decisions of pre-migrating VT task to the specific RSU of vehicular user $v$ \\
        $d_{v,e}(t)$ & Euclidean distance separating RSU $e$ and vehicular user $v$ at time slot $t$\\
        $f_v$ &  The number of GPU cycles required per unit data of vehicular user $v$  \\
        $t$ & Time slot \\
        $\xi_{v,e}(t)$ & The rendering task size of vehicular user $v$ at current serving RSU $e$\\
        $\xi_{v,e_{m}}(t)$ & The rendering task size of vehicular user $v$ at pre-migrated
RSU $e_m$\\
        $\phi_v(t)$ & Stability function of pre-migrated RSU changing\\
        $\chi_{v}(t)$ & Whether other VT tasks are simultaneously
requesting pre-migration to the same RSU\\
        $\epsilon_v(t)$ & The the bit error rate of VT task of vehicular user $v$\\

		\hline
	\end{tabular}
	\end{threeparttable}
	\label{tab:shape-functions}
\end{table}

\subsection{Vehicular Twin Migration Model}
For VT task management, the objective is to seamlessly transfer the VT task when the vehicle relocates due to changes in its position. The migration process involves relocating the VT from the current RSU to the next RSU, ensuring uninterrupted VT services. As illustrated in Fig.~\ref{fig: system}, when a vehicular user transmits a VT task request, a simultaneous decision regarding the pre-migration of the VT task must be made. Specifically, this decision involves two components: the selection by vehicular user $v$ of the RSU to which the VT tasks will be pre-migrated, denoted as $a_v(t)$, and the portion of the VT tasks to be pre-migrated, denoted as $\alpha$.

In Fig.~\ref{fig: system}, upon deciding to pre-migrate the VT task, the task is transferred to the specified RSU from the current RSU through the physical link that connects the RSUs. The bandwidth of this link, denoted as $B_{e,e_{m}}$, between the current RSU $e$ and the pre-migrated RSU $e_{m}$, determines the pre-migration latency. Hence, the pre-migration latency for VT tasks between RSUs is calculated as 

\begin{equation}
T^{m}_{v}(t)=\frac{D^{m}_{v}(t)}{B{e,e_{m}}}.
\end{equation}

\subsection{VT Task Computation Model}
While the migration model addresses the migration of VT tasks among RSUs based on the locations of vehicular users, this section delves into the VT task computation model, encompassing task processing in RSUs. The focus is on both the current serving RSU and the potential pre-migration RSU, aiming to optimize task processing and reduce latency. The considerations also involve minimizing the repetitive loading of textures in VT tasks due to frequent RSU switches, which can otherwise contribute to processing delays.

The determination of the processing conditions involves assessing whether the current serving RSU has changed since the previous time slot. If the current serving RSU remains unchanged, denoted as $I_v(t)=1$, and if there is a change, $I_v(t)=0$. Subsequently, the rendering task size on the current serving RSU is calculated as:
\begin{equation}
\xi_{v,e}(t) = D^L_v(t) - \mu I_{v}(t) D^L_v(t-1),
\end{equation}
and
\begin{equation}
D^L_v(t) = D^{task}_{v}(t) - D^{m}_{v}(t),
\end{equation}
where $D^{task}_{v}(t)$ represents the total VT task size, $D^{m}_{v}(t)$ denotes the portion of the VT task that is pre-migrated, $D^L_v(t)$ is the remaining VT task size at RSU $e$, $I_{v}(t)$ is the RSU change indicator, and $\mu$ is a coefficient accounting for the stability criterion. Consequently, the processing latency for the VT task of vehicular user $v$ at the current serving RSU $e$, from the initial queue to the completion of processing, is determined using the following equation:
\begin{equation}
T^{p}_{v,e}(t)= \frac{L_e(t) + \xi_{v,e}(t) f_v}{C_e},
\end{equation}
where current workload of RSU $e$ is represented by $L_e(t)$. The parameter $f_v$ indicates the number of GPU cycles needed per unit of data for vehicular user $v$, while $C_e$ denotes the GPU computing capacity available at RSU $e$ \cite{ren2020edge}.

To mitigate the adverse effects of frequent pre-migrated RSU switches, a stability function $\phi_v(t)$ is introduced to determine whether the pre-migrated RSU changes, defined as:
\begin{equation}
\phi_v(t) =
\begin{cases}
1, & \text{if $ a_{v}(t-1) = a_{v}(t)$,} \cr
0, & \text{otherwise.}
\end{cases}
\end{equation}

Simultaneously, the rendering task size on the pre-migration RSU, if the switching is stable, is given by:
\begin{equation}
\xi_{v,{e_{m}}}(t) = D^{m}_{v}(t) - \mu \phi_v(t)D^{m}_{v}(t-1),
\end{equation}
with the corresponding processing latency on the pre-migration RSU calculated as:
\begin{equation}
T^{p}_{v,e_{m}}(t)=\frac{L_{e_{m}}(t) +\xi_{v,e_{m}}(t) f_v}{C_{e_{m}}}.
\end{equation}

\subsection{Quality of Experience of VT Services}
In the provision of VT services, such as embodied intelligence, the vehicular user initiates a VT task request transmitted wirelessly to the RSU. The transmission of request data leads to the uplink latency $T^{u}_{v,e}(t)$. Following the submission of the VT task request, the RSU enters a waiting period for processing the VT task, resulting in a processing latency $T^{p}_{v,e}(t)$ at the current RSU. Simultaneously, based on the vehicular user's decision, a portion of the VT task is migrated from the serving RSU to a designated RSU, incurring a migration latency $T^{m}_v(t)$. Once the VT task pre-migration is completed, the time the VT task waits for processing to complete at the pre-migrated RSU generates a processing latency $T^{p}_{v,e_{m}}(t)$.

Since the processes of VT task processing and pre-migration happen simultaneously, the latency resulting from this parallel operation is calculated as
\begin{equation}
T^{p}_{v}(t)= \max \{T^{p}_{v,e}(t),T^{p}_{v,e_{m}}(t) +T^{m}_v(t) \}.\label{eq11}
\end{equation}

Upon reception of the processed VT task results, the vehicular user encounters a downlink latency $T^{d}_{v}(t)$ from the multiple RSUs. Therefore, the total latency of the VT service $T^{total}_v(t)$ can be expressed as:
\begin{equation}
T^{total}_v(t)=T^{u}_{v,e}(t)+T^{p}_{v}(t)+T^{d}_{v}(t).\label{eq18}
\end{equation}

To quantify the reliability of the VT service, the bit error rate $\epsilon_v(t)$~\cite{hieu2023enabling} is introduced to represent the reliability of the VT service, which can be calculated as:
\begin{equation}
\epsilon_v(t) = 1-\exp\left(-\sum_{v' \in V-v} \rho_{v'}(t)\chi_{v'}(t)\right),
\end{equation}
where $\rho_{v'}(t)$ represents the bit error rate contribution factor of other VT tasks:
\begin{equation}
\rho_{v'}(t) = \tau D_{v'}^m(t),
\end{equation}
and $\chi_{v'}(t)$ signifies whether other VT tasks are simultaneously requesting pre-migration to the same RSU:
\begin{equation}
\chi_{v'}(t) =
\begin{cases}
1, & \text{if simultaneous pre-migration to the same RSU,} \cr
0, & \text{otherwise.}
\end{cases}
\end{equation}

To quantify the QoE of the VT service, we propose a composite metric considering both the total latency of the VT service and the bit error rate, calculated as
\begin{equation}
QoE_v(t) = -\lambda_1\epsilon_v(t)-\lambda_2T^{total}_v(t),
\end{equation}
where $\lambda_1$ and $\lambda_2$ are weights reflecting the relative importance of bit error rate and latency in the user's QoE. This comprehensive QoE metric encapsulates both reliability and responsiveness factors. To maximize the QoE of VT services, we formulate the following optimization problem.

\subsection{Problem Formulation} 
The primary objective of this paper is to maximize user's QoE over a finite time period $T_{{max}}$. This is achieved by determining the optimal pre-migration strategies, which also ensures efficient utilization of RSU resources. The optimization problem is defined as follows:
\begin{subequations}
\begin{align} 
\max_{a} \quad &\sum_{t=1}^{T_{max}}\sum_{v=1}^{V} QoE_v(t) \label{eq:consta} \\
\text{s.t.} \quad &L_e(t) \leq L^{max}_e, \quad &&\forall e\in \mathcal{E}, \label{eq:constb} \\
&L_{e_{m}}(t) \leq L^{max}_{e_{m}}, \quad &&\forall e_{m}\in \mathcal{E}, \label{eq:constc} \\
&a_v(t) = e, \quad &&\forall v\in \mathcal{V},\forall e\in \mathcal{E}, \label{eq:constd} \\
&\alpha\in [0,1), \label{eq:conste}\\ 
&t\in {1, \ldots, T}. \label{eq:constf}\\ \nonumber
\end{align}
\end{subequations}

Constraint (\ref{eq:constb}) ensures the workload of each RSU remains within its allowable limit at all times. Similarly, constraint (\ref{eq:constc}) guarantees that the workload of pre-migration RSUs stays within their defined limits. Constraint (\ref{eq:constd}) ensures that each vehicle's VT task is assigned to only one RSU. Constraint (\ref{eq:conste}) uses the parameter $\alpha$ to ensure that the pre-migrated portion of the VT task does not surpass the task's total size. Constraint (\ref{eq:constf}) sets the framework for the optimization problem is addressed within a specific time horizon $T_{{max}}$. Since the VT task migration optimization problem is NP-hard due to its Markov property~\cite{chen2024aiot}, we formulate the problem as a POMDP and solve it using MADRL.

\section{Efficient Trajectory Generation for MADRL} \label{Gen}
In this section, we introduce EST-TG, an efficient spatio-temporal trajectory generation algorithm. To implement an efficient trajectory generation algorithm, the available data needs to be analyzed as well as preprocessed. Therefore, we first present the preparation of the dataset as well as the data preprocessing. Then, the processed data is leveraged to analyze it to extract key features. Finally, the details of the proposed trajectory generation algorithm are presented.
\begin{figure*}
    \centering
    \includegraphics[width=1\linewidth]{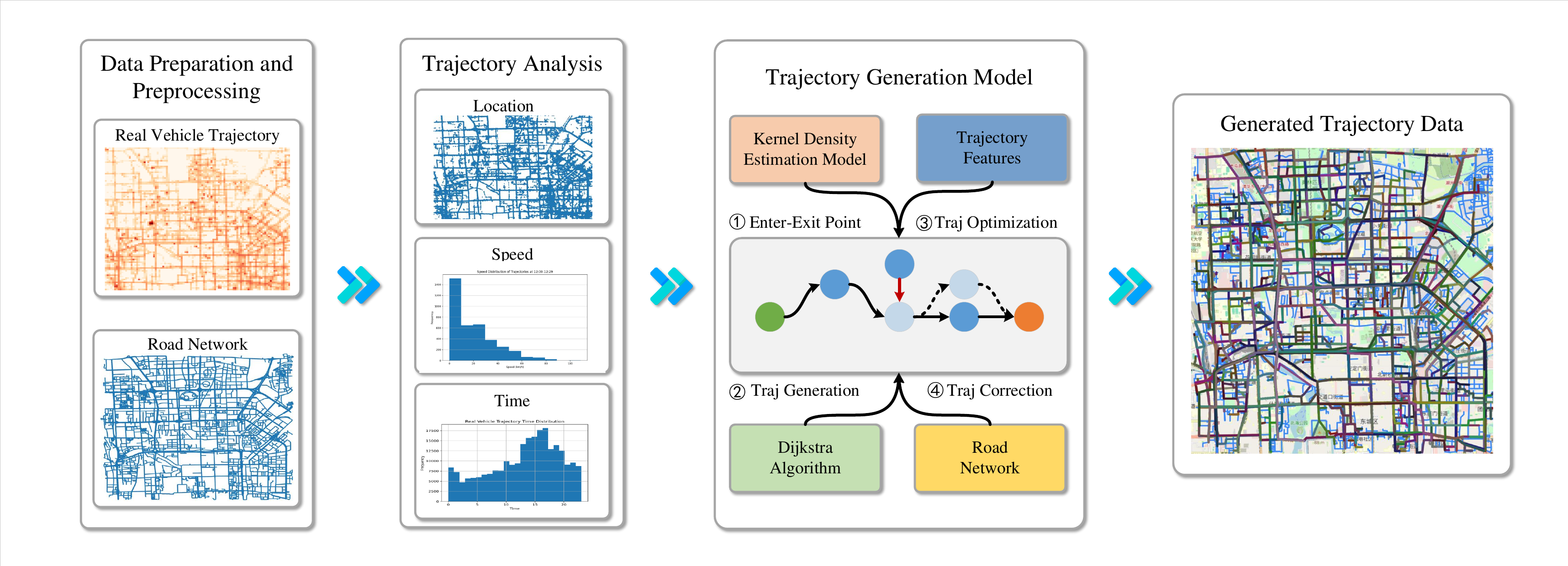}
    \caption{Architecture of the EST-TG algorithm for spatio-temporal trajectory generation.}
    \label{fig: gen}
\end{figure*}

\subsection{Data Preparation and Preprocessing}
To enable further analysis of the spatio-temporal motion patterns of vehicle trajectories, we first load the vehicle trajectory dataset and segment the vehicle trajectories. Specifically, for each vehicle, its anomaly data will be cleared and the trajectories will be segmented by successive periods. Let $Traj^{ori}_{v} = \{P_v^1,\dots,P_v^t,\dots,P_v^T | P_v^t = (x_v^t,y_v^t)\}$ represents the origin trajectory of vehicular user $v$ with GPS sample point $P_v^t = (x_v^t,y_v^t)$ at time step $t$. The original vehicle trajectory included GPS measurement noise and did not consider the geographical correlation between different locations \cite{ji2016comparison}. Additionally, during the trajectory generation process, the resulting samples might be unrealistic. For instance, GPS coordinates could fall outside of roadways or result in disjointed route segments. To address these issues, we utilize the data of the road network to correct the off-road GPS coordinates~\cite{murphy2019map}. Specifically, we load the data of the road network and convert GPS location position sequences into road-connected sequences, denoted as $Traj^{ori-rc}_{v} = \{l_v^1,\dots, l_v^t,\dots, l_v^T | l_v^t = f(RN, P_v^t) \}$, where $RN$ is the road network in the region corresponding to the original vehicle trajectory, and $f(\cdot)$ is a matching algorithm~\cite{zhu2022knowledge} that maps GPS position points to the nearest road network.
\subsection{Trajectory Data Analysis and KDE Application}
After completing the preparation and pre-processing of vehicle trajectory data and road network data, we analyze the temporal and spatial distribution characteristics of the vehicle trajectory data to generate trajectories whose temporal and spatial distributions conform to the real vehicle movement characteristics. 

We adopt the Kernel Density Estimation (KDE) model to analyze the spatial distribution of vehicle trajectories due to its efficiency, flexibility, accuracy, and smoothness in handling complex data. KDE offers a more computationally efficient solution by applying a kernel function to each data point, which avoids the need for binning as in histogram estimation, ensuring continuous density estimates without discontinuities, which is crucial for analyzing spatially continuous data like vehicle trajectories. Moreover, KDE allows the selection of different kernel functions and bandwidths, providing adaptability to different data characteristics. Compared to methods like k-NN, which can be computationally intensive due to distance calculations and sensitive to outliers, KDE uses localized weighting, making it both more efficient and robust in capturing the underlying patterns of vehicle movement. The KDE model can be expressed as:
\begin{equation}
    \hat{f_h}(x) = \frac{1}{n} \sum_{i=1}^{n} K_h(x - x_i),
\end{equation}
where \( \hat{f_h}(x) \) is the estimated density at point \( x \), \( n \) is the number of data points, \( K_h \) is the kernel function with kernel bandwidth \( h \), and \( x_i \) represents the data points.

For vehicle trajectory data, we consider each trajectory point as an observation and distribute it over the entire study area. By applying the KDE model, we can generate spatial density maps of the vehicle trajectory data, showing the degree of aggregation of vehicles at different geographical locations. These density maps can help us to capture traffic congestion areas, hotspot roads, and possible traffic flow patterns, making the generated trajectory data more consistent with the real-world vehicle data distribution.

Next, we conduct analyses of the speed distribution in the vehicle trajectory dataset and the time distribution within 24 hours. We first analyze the speed distribution in the vehicle trajectory data set. By extracting the speed information in the trajectory data, we can get the actual traveling speeds of vehicles at different locations and time points. This can help us gain insights into the trend of vehicle traveling speed and the speed distribution law. Next, we analyze the time distribution of the vehicle trajectory data over 24 hours. By labeling and classifying the trajectory data by time, we can get the number or density of vehicle trajectories in each period. This can help us identify traffic peaks and troughs, as well as the changing patterns of traffic flow during different periods.




\subsection{Trajectory Generation Algorithm}

In this section, we provide a detailed description of our proposed trajectory generation algorithm. Illustrated in Fig.~\ref{fig: gen}, our algorithm comprises four key phases: entry-exit point generation, trajectory generation, trajectory optimization, and trajectory correction. To execute each phase effectively, we leverage the KDE model, trajectory features, Dijkstra's algorithm, and road network data.

\subsubsection{Entry-Exit Point Generation} 
The initial phase of our algorithm involves the identification and generation of entry and exit points for trajectories. To generate trajectory data that matches the real-world traffic conditions at various times, our proposed algorithm employs different KDE models $\hat{f_h}(x)$ for different times of the day. These models are used to generate entry and exit points, with the number of points determined by the distribution density of the original dataset's trajectory points at each time, identifying areas of high traffic flow as potential entry and exit points. Let \( P_{entry} \) and \( P_{exit} \) denote the sets of entry and exit points, respectively, and \( P_v^1 \in P_{entry} \) and \( P_v^T \in P_{exit} \) represent individual entry and exit points generated by the KDE model. 

\subsubsection{Trajectory Generation}
Following the identification of entry and exit points, the trajectory generation phase, as the core phase in EST-TG, constructs feasible trajectories connecting these points. Leveraging edges and nodes from road network data, we employ a graph-based algorithm to generate candidate trajectories. The Dijkstra's algorithm can be formulated as follows:
\begin{equation}
d(v) = \min(d(v), d(u) + w(u, v)),
\end{equation}
where \( d(v) \) represents the shortest distance from the source node to node \( v \), \( d(u) \) is the shortest distance from the source node to node \( u \), and \( w(u, v) \) denotes the weight of the edge connecting nodes \( u \) and \( v \). By incorporating Dijkstra's algorithm to generate vehicle trajectories based on the road network data, we ensure the efficiency and optimality of trajectory paths within the road network.

\subsubsection{Trajectory Optimization}
After generating the candidate trajectories, the optimization phase improves these trajectories to enhance their efficiency and make them more compatible with real-world traffic conditions. First, we optimize the generated trajectory by incorporating real-world speed data. For each consecutive trajectory point \( TP_i = (x_i, y_i) \) and \( TP_{i+1} = (x_{i+1}, y_{i+1}) \), we calculate the corresponding traveling time \( t_{i+1} \) based on the vehicle speed \( v_i \) at point \( TP_i \), derived from the speed distribution data. The travel time is computed as:
\begin{equation}
t_{i+1} = t_i + \frac{d(TP_i, TP_{i+1})}{v_i},
\end{equation}
where \( d(TP_i, TP_{i+1}) \) represents the Euclidean distance between the two points.

Next, we apply a uniform time-based interpolation method to generate additional points between \( TP_i \) and \( TP_{i+1} \) to ensure that the trajectory follows a consistent time interval. For each time step \( t_j = t_i + j \Delta t \) where \( t_j < t_{i+1} \) and interpolation time interval \( \Delta t > 0 \), we interpolate new points \( TP_j = (x_j, y_j) \) using the following formulas based on linear interpolation:

\begin{equation}
\begin{aligned}
x_j &= x_i + \frac{t_j - t_i}{t_{i+1} - t_i}(x_{i+1} - x_i), \\
y_j &= y_i + \frac{t_j - t_i}{t_{i+1} - t_i}(y_{i+1} - y_i),
\end{aligned}
\end{equation}
where \( t_j = t_i + j \Delta t \) represents the time at which the interpolated point \( TP_j \) is generated. By using this method, the generated trajectory reflects realistic vehicle movement in both time and space and can adapt to specific time interval requirements.

\subsubsection{Trajectory Correction}

The final phase of our algorithm is trajectory correction, which deals with any discrepancies or inaccuracies in the generated trajectories. Due to the trajectory densification algorithm used in the trajectory optimization phase, some of the trajectories deviate from the actual road network constraints. Therefore, we use the previously proposed trajectory correction algorithm to adjust the generated trajectories to ensure that the generated trajectories comply with the road network constraints.

\section{Multi-agent Split Reinforcement Learning for Pre-migration Decisions}\label{madrl}
In the vehicular metaverse, sufficient training data is critical for model training, e.g., the model training in MADRL. However, the costly and scarce long-distance data pose a challenge. By simulating realistic traffic patterns, the EST-TG algorithm provides enriched environments that aid MADRL in modeling complex interactions among vehicles and RSUs that are crucial for VT migrations within vehicular metaverses. 
\subsection{POMDP for VT Task Pre-migration}
In vehicular metaverses, the pre-migration decisions of vehicular users are influenced by various factors, such as user mobility, RSU density, RSU workloads, VT task requests from other users, and the pre-migration choices made by other vehicular users. When vehicular user $v$ possesses full information of the decision process, it can determine the optimal  VT tasks pre-migration strategy. Nevertheless, obtaining all the necessary information is impractical. Additionally, the workload of an RSU is dynamically influenced by VT tasks from other users during each time slot. To address the challenge of making optimal decisions with partially observable information, we model the VT task migration problem as a POMDP, defined as follows.

\begin{figure*}[t]  	
\centerline{\includegraphics[width=1\linewidth]{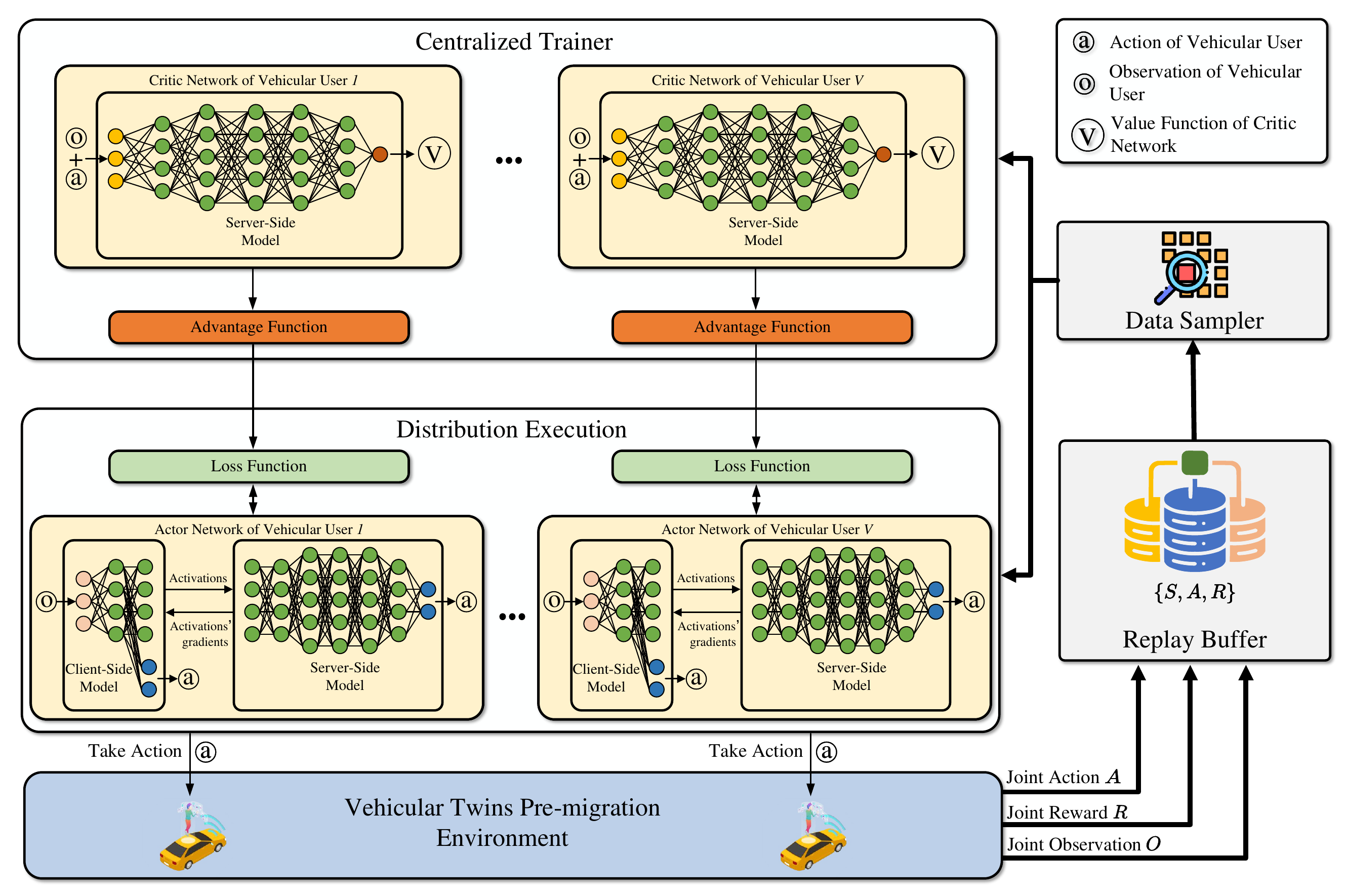}}  	
\caption{Design of the MSRL Algorithm for Pre-migration of VT Tasks.} \label{fig: MSRL}  
\end{figure*}

\begin{itemize}
    \item \textbf{Observation:} In the vehicular metaverse, vehicular users monitor data about RSUs and other users in real time. The collective observations from all users in the environment are denoted as $\mathcal{O}$, represented by $\mathcal{O} = \{ o_1, \dots,o_v, \dots, o_V\}$. For a specific user $v$, the observation at time slot $t$ is described as follows:
    \begin{equation}
        o_v(t)=[a_v(t),L_{\mathcal{E}}(t),\epsilon_v(t),\phi_v(t),\chi_{v}(t),T^{total}(t)],\label{obs}
    \end{equation}
    where $a_v(t)$ represents the pre-migration decision of vehicular user $v$, $L_{\mathcal{E}}(t)$ denotes the available workload of RSUs at time slot $t$, signifies the bit error rate of vehicular user $v$'s VT task, $\phi_v(t)$ represents the stability function of pre-migrated RSU switching, $\chi_{v}(t)$ indicates whether tasks of other VTs are simultaneously requesting pre-migration to the same RSU, and $T^{total}_v(t)$ is the total VT services latency of vehicular user $v$ at time slot $t$. 
    \item \textbf{Action:} Each time slot, vehicular user $v$ can execute the pre-migration decision, i.e., $a_v(t)$ with discrete action space that is vehicular user $v'$s selection of the RSU to which to pre-migrate VT tasks at time slot $t$. 

    \item \textbf{Reward:} Each time slot, a vehicular user receives an observation $o_v^t$ and performs action $a_v(t)$ based on the pre-migration policy. The environment returns a reward during the interaction, defined as:
    \begin{equation}
    r_v(t)= -T^{total}_v(t), \label{eqrew}
    \end{equation}
    where $T^{total}_v(t)$ is the total latency for VT service latency of vehicular user $v$ at time slot $t$.
\end{itemize}

\subsection{MSRL Algorithm Design}
Coordination among agents is vital for optimizing complex tasks in multi-agent systems. This is particularly relevant in vehicular metaverse environments, where multiple vehicular users navigate and interact. In this paper, we propose a Multi-Agent Split Reinforcement Learning (MSRL) scheme to enhance the decision-making process for VT pre-migration. The MSRL algorithm is designed to achieve efficient collaboration among multiple learning agents. It integrates centralized training with decentralized execution. In this framework, each agent operates with an independent policy network while sharing a unified value network. This design enables all agents to access a global valuation, thereby improving collaboration and overall performance.

One noteworthy feature of the MSRL model is its incorporation of a split architecture. The actor network employs a split architecture with two distinct paths: one for a small neural network on the local side and another for a large neural network on the server side. Both the small and large networks produce intermediate outputs, and the final action logits are obtained from the output layers. The output is further processed through a softmax layer to obtain a probability distribution over the action space. As shown in Fig.~\ref{fig: MSRL}, for vehicular user $v$, its actor network consists of a client-side model and a server-side model. For the client-side model, its input is $o_v^t$, the observation of vehicular user $v$ at time slot $t$, then action distributions are generated to further obtain the action $a_v(t)$.  For the server-side model, its inputs are the intermediate outputs of the client-side model network, and its outputs are the same as those of the client-side model, both of which are $a_v(t)$.
The decision to use either only the client-side model or both of the models is dynamically determined based on the entropy of action distribution on the client-side model, where the mean entropy of the action distribution is monitored. Large networks on the server-side model are activated if the mean entropy exceeds a critical value or fluctuates constantly around the critical value, indicating that the client-side model is not capable of adapting to the complexity of the environment. To achieve adaptive switching, we set up two mechanisms, namely, the dynamic threshold adjustment mechanism and the switching buffer mechanism.

\subsubsection{Dynamic Threshold Adjustment Mechanism}
As the DRL algorithm continues to be trained, the distribution of actions in the output of the policy network will stabilize, causing the entropy value of the output to decrease. Therefore, the threshold needs to be varied accordingly to ensure that it remains relevant to the current state of the model. The dynamic threshold adjustment mechanism is expressed as
\begin{equation}
Thr^s = Thr^0 + \sum_{s=0}^S Lar^s Ch,\label{dam}
\end{equation}
where $Thr^s$ denotes the value of the threshold at the $s$-th algorithmic call to the actor, $Thr^0$ denotes the initial threshold, $S$ denotes the total number of current algorithmic calls to the actor, $Lar$ is a binary value that is $1$ when the actor calls the server-side model and $Ch$ denotes the threshold changing factor. The decision to use the large network is based on the mean entropy that is calculated over a fixed number of recent observations. The switching logic is determined by whether the mean entropy surpasses the threshold or exhibits fluctuations.

When the entropy value reflects that the vehicle-side model makes it difficult to cope with the complex environment, the system automatically activates the edge server-side model. This mechanism ensures that more powerful edge computing resources can be utilized in a timely manner to improve the quality of decision-making when the vehicle decision-making capability is insufficient, while unnecessary computational overhead is avoided when the environment is simpler, thus improving the efficiency of resource utilization.

\subsubsection{Switching Buffer Mechanism}
During model switching, frequent switching between client-side and server-side models can lead to an unstable training process, thus affecting the convergence speed of the model. Therefore, we will calculate the mean entropy value over a certain period to judge the network selection. During the training process, the selected network would be trained. When the mean entropy value within the period is greater than a threshold, the algorithm transfers the intermediate results from the small network of the client-side model to the large network of the server-side model to generate the agent's actions, and both of the networks would be trained. Otherwise, only the small network of the client-side model would be trained and used to generate the agent's actions. In addition, when the algorithm switches frequently between the small and large networks, in order to maintain the stability of the algorithm training, the small network of the client-side model and the large network of the server-side model will be trained at the same time, and the large network of the server-side model will be used to generate the action of the agent, which avoids unnecessary switching due to short-term fluctuations, thus maintaining the stability of the algorithm training process.

For each agent $v$, the action policy is defined as $\pi_{\theta_v}[a_v(t)|o_v^t]$. Both the client-side and server-side models are updated independently. To mitigate instabilities from significant discrepancies in policy network updates, the policy constrains its updates by minimizing the respective clipping surrogate targets. This can be expressed as

\begin{equation}
L^{clip}(\theta_v) = \mathbb{E}\{F(\beta_{\theta_v}^i,A_v[o_i,a_i])\},\label{ld}
\end{equation}
\begin{equation}
F(\beta_{\theta_v}^i,A_v[o_i,a_i])=min[\beta_{\theta_v}^t A_v,{f}_{clip}[\beta_{\theta_v}^t,\epsilon] A_v],\label{fund}
\end{equation}
where ${f}_{clip}[\beta_{\theta_v}^t,\epsilon] A_v]$ is the clip function, with $\epsilon$ in the range [0,1] as the clipping parameter. The ratio of new to old policies is represented by $\beta_{\theta_v}^t$, defined as
\begin{equation}
\beta_{\theta_v}^t=\frac{\pi_{\theta_v} [a_v(t)|o_v^t]}{\pi_{\theta_v}^{old} [a_v(t)|o_v^t]},\label{db} 
\end{equation}
where $\pi_{\theta_v}^{old} [a_v(t)|o_v^t]$ is the old policy and $\pi_{\theta_v} [a_v(t)|o_v^t]$ is the new policy. The joint advantage function $A_v[o^t,a(t)]$ evaluates the advantage of action $a(t)$ over the average action in state $o^t$ \cite{DelinGuo2020MAPPO}, calculated as
\begin{equation}
A_v[o^t,a(t)]=\hat{Q}_v[o^t, a(t)]-b[o^t, a_{-v}(t)],\label{advd} 
\end{equation}
where $\hat{Q}_{v}[o^t, a(t)]$ is the estimate of the action-value function, computed as 
\begin{equation}
\begin{aligned}
\hat{Q}_v[o^t, a(t)]&=Q_{\bar{\omega}_v}[o^t, a(t)] \\
& \quad +\delta_{t}+(\gamma \lambda) \delta_{t+1}+\cdots+(\gamma \lambda)^{T} \delta_{T},\label{hatqd}
\end{aligned}
\end{equation}
where $Q_{\bar{\omega}_v}[o^t, a(t)]$ is the centralized critic, $\delta_{t}$ denotes the temporal difference (TD) error, $\gamma$ is the discount factor, and $\lambda$ is the decay factor applied to the TD error. $b[o^t, a_{-v}(t)]$ and $b[o^t, a^c_{-v}(t)]$ act as counterfactual baselines, where $a_{-v}(t)$ and $a^c_{-v}(t)$ denote the joint actions of all agents except agent $v$.

Algorithm \ref{alg:algorithm} presents the pseudo-code for the proposed MSRL algorithm. Initially, we set up the actor $\pi_{\theta_v}$, $\pi_{\theta_v}^{old}$, critic $Q_{\omega_v}$, $Q_{\bar{\omega}_v}$. Additionally, we initialize the training environment $env$ and the replay buffer $D$.

As depicted in Algorithm \ref{alg:algorithm}, during the training phase, each agent's actor adaptively switches between the client-side model and the server-side model based on the dynamic threshold adjustment mechanism and the switching buffer mechanism, generating action $a_v(t)$  based on their observations $o^t_v$, output entropy, and model utilize stability. The agents then receive the subsequent state $o^{t+1}$ and the reward $r^t$ from $Env$. Following this, each agent $v$ obtain its trajectory $\tau_v$ and compute ${\{\hat{Q}_v[o^t,a(t)]\}}_{t=1}^{T}$, and ${\{A_v[o^t,a(t)]\}}_{t=1}^{T}$. The experiences ${\{o,a,\{\hat{Q}(o,a)\},\{A(o,a)\},r\}}$ is stored in the $D$ after the environment transition is complete. During each training epoch $k$, data order is randomized to break sample correlations, enhancing stability. Subsequently, $\theta_v$ and $\omega_v$ are updated using gradient $\Delta\theta_v$ and $\Delta\omega_v$ with $G$ mini-batch of data samples drawn from $D$. After $K$ training epochs, each agent's $\theta_v^{d(old)}$ and $\bar{\omega}_v$ are updated to the new $\theta_v^{d}$ and ${\omega}_v$.

\begin{algorithm}[H]
\caption{MSRL Algorithm for VT Task Pre-migration}
\label{alg:algorithm}
\begin{algorithmic}[1] 
\STATE Initialize  actor $\pi_{\theta_v}$, $\pi_{\theta_v}^{old}$, critic $Q_{\omega_v}$; training environment $env$, and replay buffer $D$;
\FOR {$E$ in $1,\dots,eps$}
\FOR{$t$ in $1,\dots,T$}
\STATE Insert generated vehicle trajectories into $env$;\\
\FOR{Agent $v=1,2,\dots,V$}
\STATE Input observation $o_v^t$ into client-side model of actor $\pi_{\theta_v}^{old}$ then obtain action $a_v(t)$ and entropy;\\
\STATE Apply the dynamic threshold adjustment mechanism and the switching buffer mechanism to the entropy to decide whether to activate the server-side model;\\
\IF{Use server-side model}
\item Transfer the intermediate results to the server-side model of actor $\pi_{\theta_v}^{old}$ to generate the action $a_v(t)$ and overwrite the original action;\\
\ENDIF
\STATE Execute action $a_v(t)$ to obtain the next state $o^{t+1}$ and reward $r^t$;
\ENDFOR
\ENDFOR
\STATE Each agent $v$ collects a trajectory $\tau_v=\{o_v^t,a_v^t,r_v^t\}_{t=1}^{T}$;
\STATE Compute ${\{\hat{Q}_v[o^t,a(t)]\}}_{t=1}^{T}$ according to Eq.~(\ref{hatqd});
\STATE Compute ${\{A_v[o^t,a(t)]\}}_{t=1}^{T}$ according to Eq~(\ref{advd});
\STATE Store ${\{o,a,\{\hat{Q}(o,a)\},\{A(o,a)\},r\}}$ into $D$;

\FOR{Epoch $k$ in $1,...,K$}
\STATE Randomized the data order in $D$
\FOR{$j$ in $0,1,2,...,{\frac{T}{G}}-1$}
\STATE Sample $G$ mini-batch of data from $D$;
\FOR{v=1,2,...,V}
\STATE $\Delta\theta_v=\frac{1}{G}\sum_{i=1}^G\{{\nabla}_{\theta_v}F(\beta_{\theta_v}^i,A_v[o_i,a_i])\}$;
\STATE $\Delta\omega_v=\frac{1}{G}\sum_{i=1}^G\{{\nabla}_{\omega_v}[\hat{Q}_v(o_i,a_i)-Q_{\omega_v}(o_i,a_i)]^2\}$;
\ENDFOR
\STATE Update $\theta_v$, $\omega_v$ using $\Delta\theta_v$, $\Delta\omega_v$;
\ENDFOR
\ENDFOR
\STATE Set $\theta_v^{old}=\theta_v $, $\overline{\omega}_v=\omega_v$.
\ENDFOR
\end{algorithmic}
\end{algorithm}


For the MSRL algorithm, the actor-network size of the client-side model is $n$, and the actor-network size of the server-side model is $n'$. When the actor only uses the client-side model, the computation complexity of the actor-network is $O(n)$; otherwise, the computation complexity of the actor-network is $O(n + n')$. Therefore, the total computation complexity of the MSRL algorithm is $O[epsTV(N + \frac{K}{D})]$, where $N$ represents the computation complexity of the actor-network for all agents~\cite{chen2024aiot}.

\section{Numerical Results}\label{result}
Initially, we analyze the similarity between the trajectories generated by the trajectory generation algorithm and the real trajectories in terms of spatio-temporal distribution. Next, we analyze the convergence performance of the proposed MSRL algorithm. Finally, we compare the performance of MSRL with other baselines across different scenarios.
\subsection{Parameter Settings}
We use a real-world vehicle movement dataset~\cite{yuan2010t} and a real-world traffic road network dataset to measure the performance of our proposed algorithm. The \textit{T-Drive} dataset contains a real GPS trajectory dataset of Beijing cabs from February 2 to 8, 2008, which is sampled once per minute. In addition, the Beijing traffic network dataset is extracted from \textit{Open Street Map}. To better generate vehicle trajectories, we only extract the major roads.

In the EST-TG algorithm, we use the KDE model with a Gaussian kernel to estimate the spatial distribution of vehicle movements. The bandwidth parameter $h$ is set to $0.05$, and the interpolation time interval $\Delta t$ is set to $30$ seconds.

In the MSRL algorithm, we set the learning rate to $10^{-3}$ and the size of the discount factor to $0.95$. Actor's network size is set to $[8, 16, 16, 32, 16]$, and the splitting point is between the second-layer network and the third-layer network. The size of the critic network is set to $[32,32]$. The entropy threshold for switching between client-side and server-side models is initially set to $0.7$ and dynamically adjusted during training.
\subsection{Trajectory Generation}
In this section, we present a detailed analysis of the experimental results obtained from trajectory generation. 

\subsubsection{Spatial Distribution Analysis}
\begin{figure}
    \centering
    \includegraphics[width=1\linewidth]{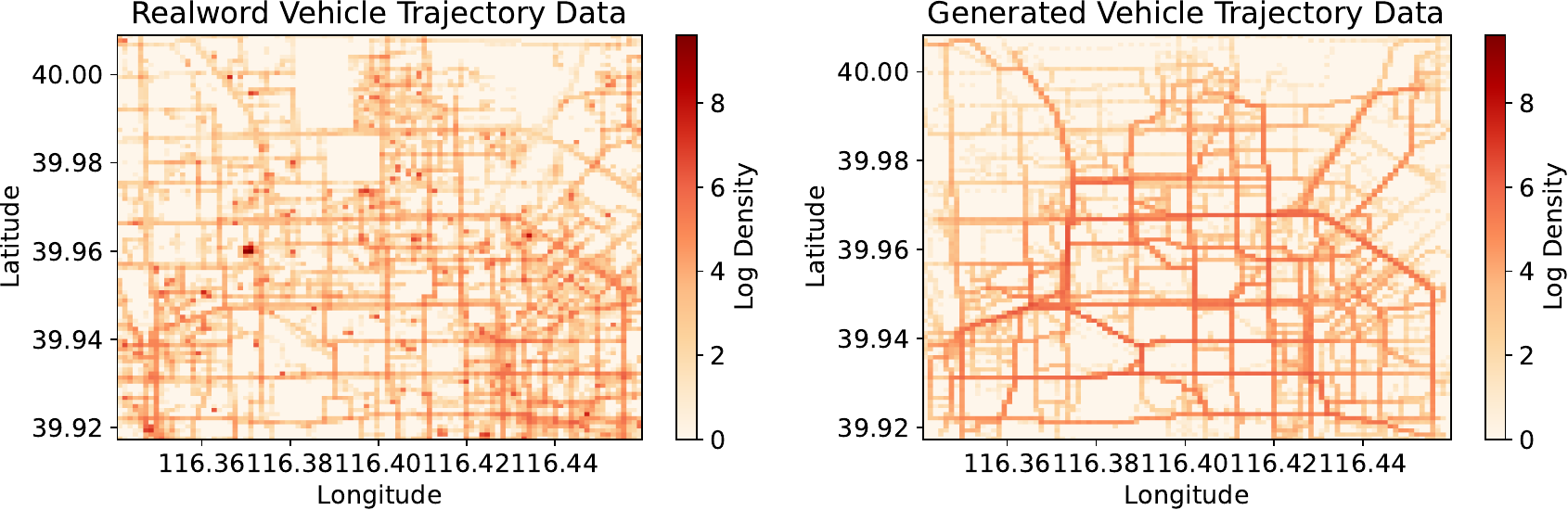}
    \caption{Real vehicle trajectory spatial distribution data v.s. generated vehicle trajectory spatial distribution data.}
    \label{fig:heat map}
\end{figure}

Figure \ref{fig:heat map} illustrates a comparison between the spatial distributions of real vehicle trajectories and the generated trajectories. The heat map visualization demonstrates that the generated trajectories closely resemble the spatial patterns observed in actual vehicle movements. Notably, there is a discernible correspondence between high-traffic regions in both datasets, indicating the effectiveness of the proposed trajectory generation algorithm in capturing spatial dynamics accurately.
\begin{figure}
    \centering
    \includegraphics[width=1\linewidth]{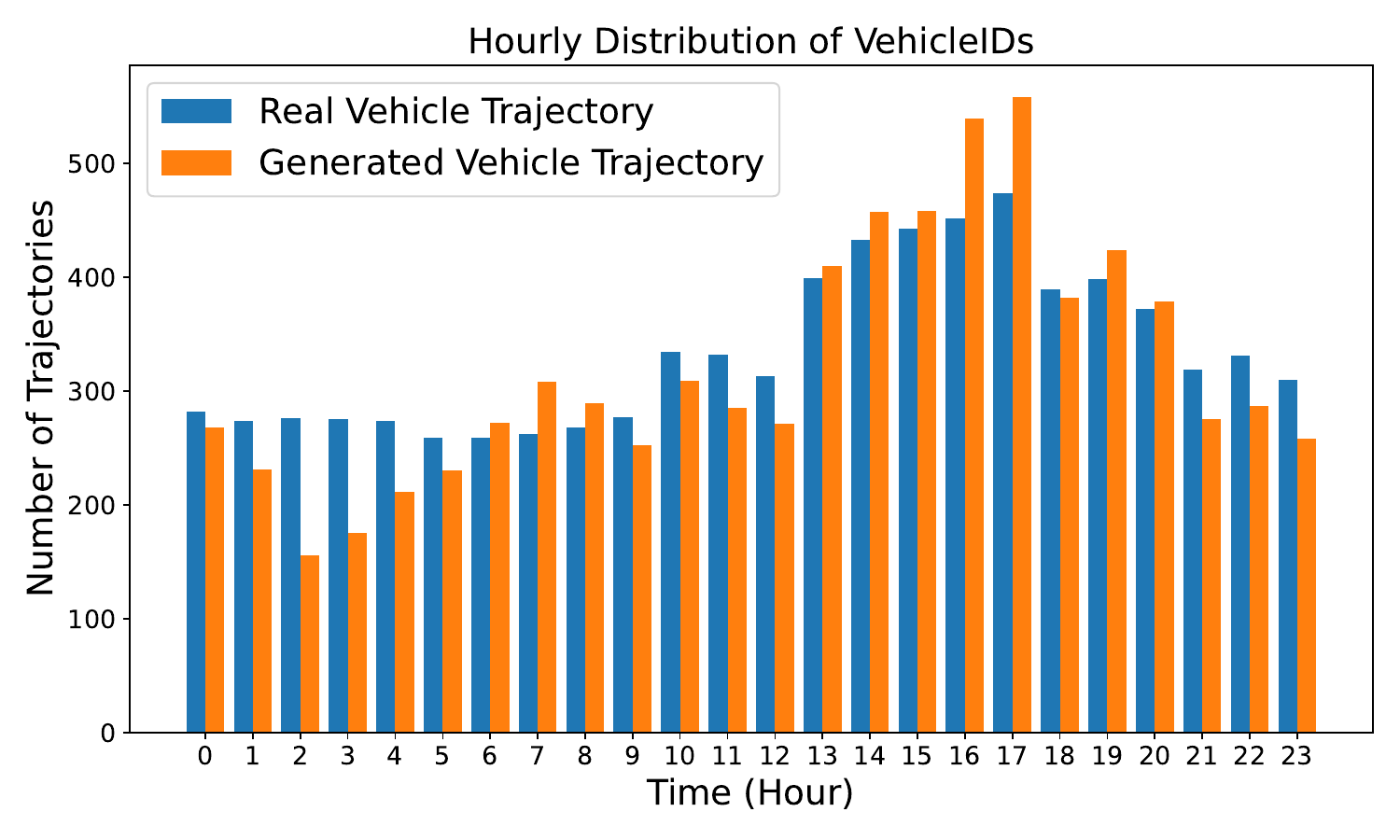}
    \caption{Comparison of real vehicle time distribution in a day against the generated vehicle time distribution.}
    \label{fig:time-dist}
\end{figure}
\subsubsection{Temporal Distribution Comparison}
Figure \ref{fig:time-dist} presents a comparative analysis of the temporal distribution of real vehicle movements throughout the day and the generated counterparts. This comparison reveals the temporal fidelity of our generated trajectories, as they mirror the hourly distribution patterns observed in real-world vehicular activity. Despite inherent variations, such as peak hours and lulls in traffic, our scheme successfully replicates the temporal characteristics of genuine vehicle movements, thereby enhancing the realism of generated trajectories.
\begin{figure}
    \centering
    \includegraphics[width=1\linewidth]{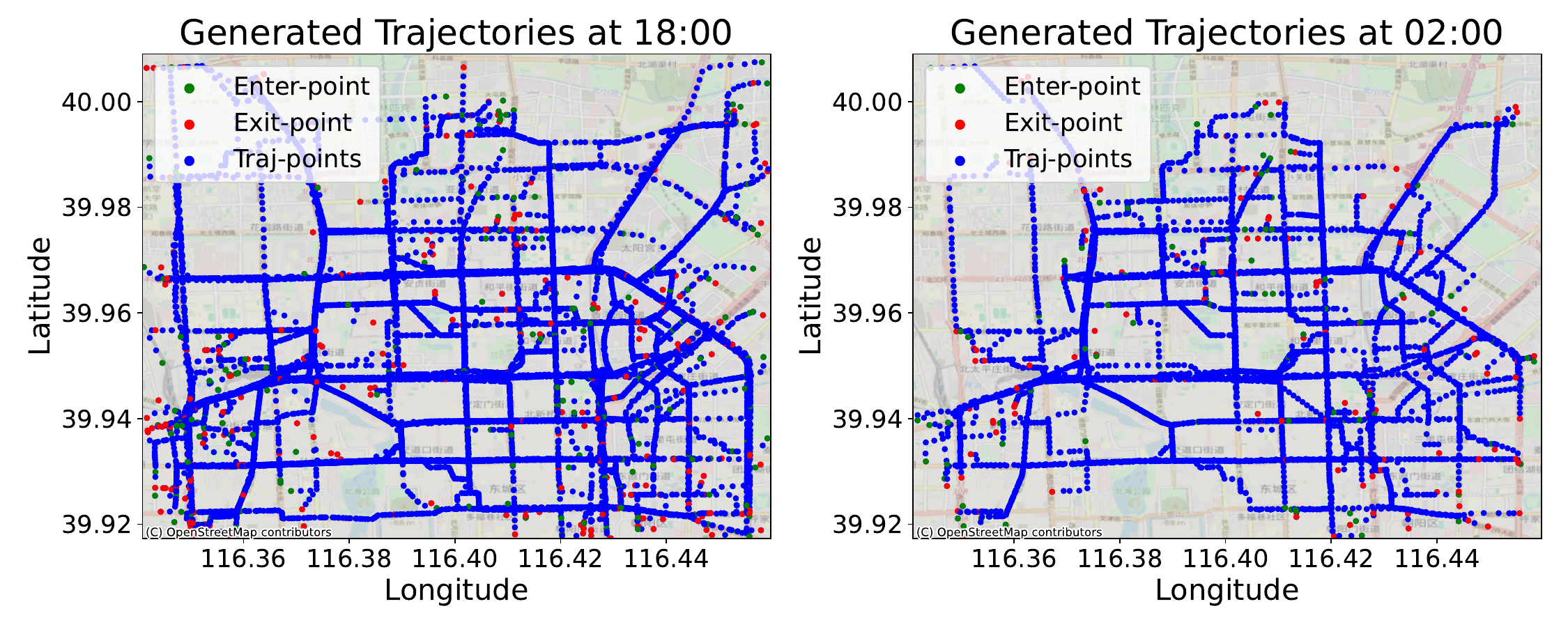}
    \caption{Rush hour vehicle trajectory map vs. early morning vehicle trajectory map.}
    \label{fig:visual-map}
\end{figure}
\subsubsection{Trajectory Mapping and Time Differential Visualization}
The visualization depicted in Fig. \ref{fig:visual-map} contrasts rush hour vehicle trajectories with those observed during the early morning hours. Simulation results show that the number of generated trajectories is higher in the rush hour and fewer in the early morning. Moreover, the generated trajectories are slower near the city center and faster near the periphery of the city. Since the 
 proposed scheme uses road network matching algorithm, all the generated vehicle trajectories have good road network matching. 

\subsection{Convergence Analysis}
To evaluate the efficiency of our proposed MSRL algorithm, i.e., \textit{Split Model}, and to show the effectiveness of the proposed algorithm in reducing the computational parameters, we designed several baselines for comparison, including \textit{Local-Edge Model}, \textit{Local-Model}, \textit{Full Migration}, and \textit{Random Migration}. Among the baseline algorithms, \textit{Local-Edge Model} denotes the actions generated by the agent's reasoning using the edge server-side model, \textit{Local Model} denotes the actions generated by the agent's reasoning using the client-side model, \textit{Full Migration} denotes the agent always migrates all tasks to the nearest server, and \textit{Random Migration} indicates that the agent randomly pre-migrates some tasks to nearby servers.

\begin{figure}
    \centering
    \includegraphics[width=1\linewidth]{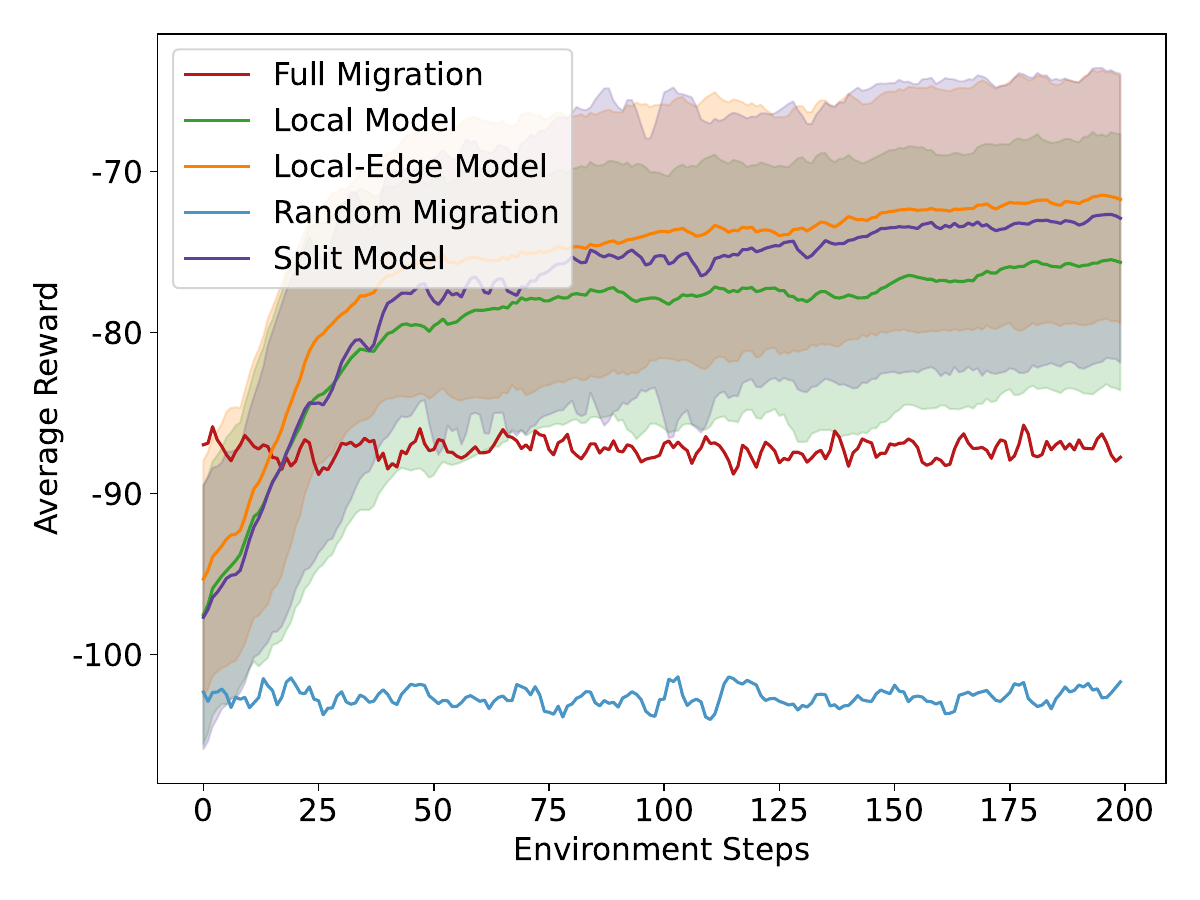}
    \caption{Average reward versus number of environment steps.}
    \label{fig:rew-curve}
\end{figure}

Fig.~\ref{fig:rew-curve} illustrates the curve of the average reward as a function of the number of steps in the environment. As shown in Fig.~\ref{fig:rew-curve}, \textit{Split Model} outperforms \textit{Local-Model}, \textit{Full Migration}, and \textit{Random Migration} by 6\%, 10\%, and 29\%, respectively. It is worth noting that \textit{Split Model} has a higher average reward and converges faster compared to \textit{Local-Model}, illustrating the effective learning and decision-making capabilities of the proposed  MSRL algorithm.

\begin{figure}
    \centering
    \includegraphics[width=1\linewidth]{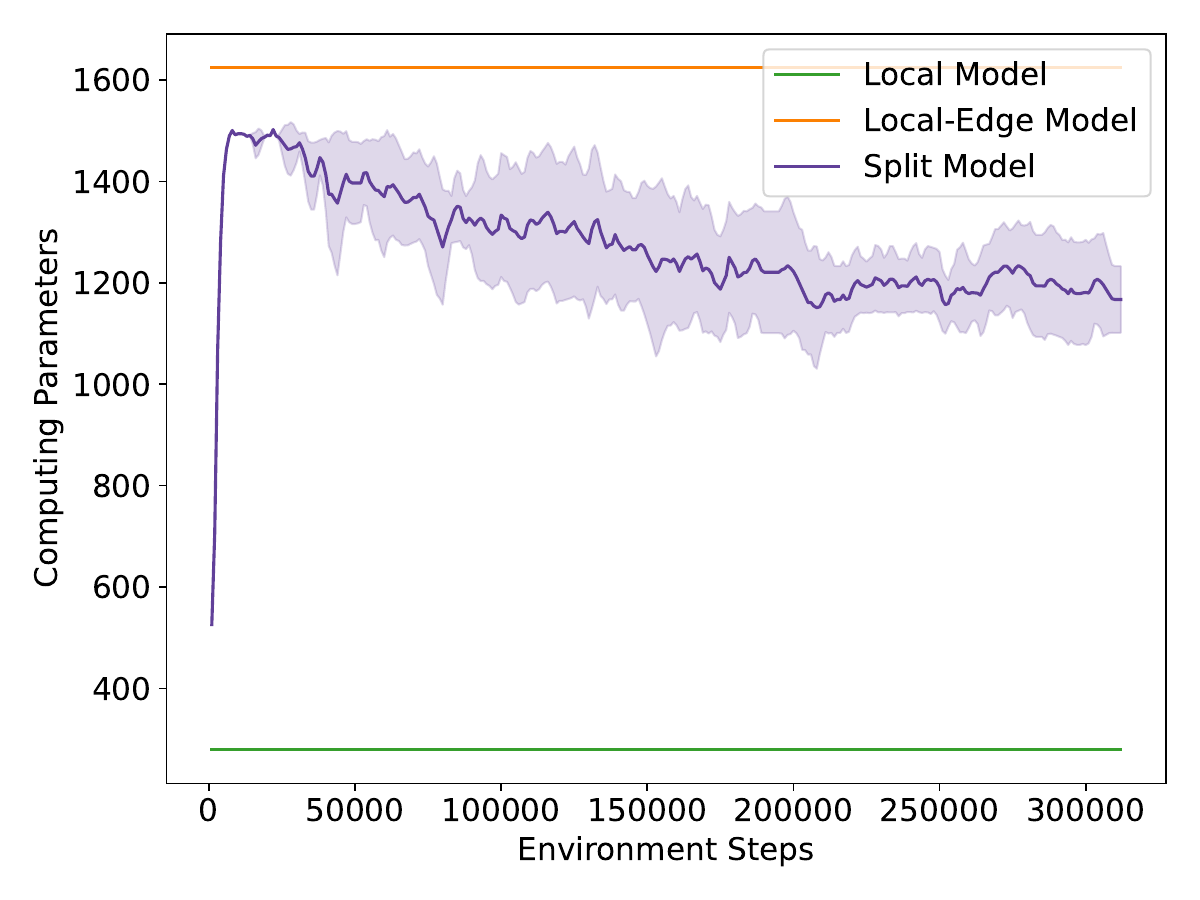}
    \caption{Computing Parameters versus number of environment steps.}
    \label{fig:param-curve}
\end{figure}

In addition, as shown in Fig.~\ref{fig:rew-curve} and Fig.~\ref{fig:param-curve}, the number of computational parameters of \textit{Local-Model} is less, which is about 200, and its computational load is less and the representation ability of the model is weaker, so its average reward is lower. Meanwhile, \textit{Local-Edge Model} has a higher number of computational parameters, about 1600, which indicates a heavier computational load, but its model representation ability is stronger, so it converges faster and has a higher average reward. In contrast, the number of parameters for \textit{Split Model} decreases as the algorithm continues to be trained, eventually converging to approximately 1200. Despite the 25\% reduction in the number of computed parameters, \textit{Split Model}'s performance only slightly lags behind that of \textit{Local-Edge Model} by about 2\% and outperforms \textit{Local-Model} in terms of both average reward and convergence speed. The reason is that the distribution of actions in the output of the policy network stabilizes as the algorithm continues to be trained, causing the entropy value of the output to decrease, therefore increasing the probability that the algorithm is calling \textit{Local-Model}.



\subsection{Performance Evaluation in Different Scenarios}
\begin{figure}
    \centering
    \includegraphics[width=1\linewidth]{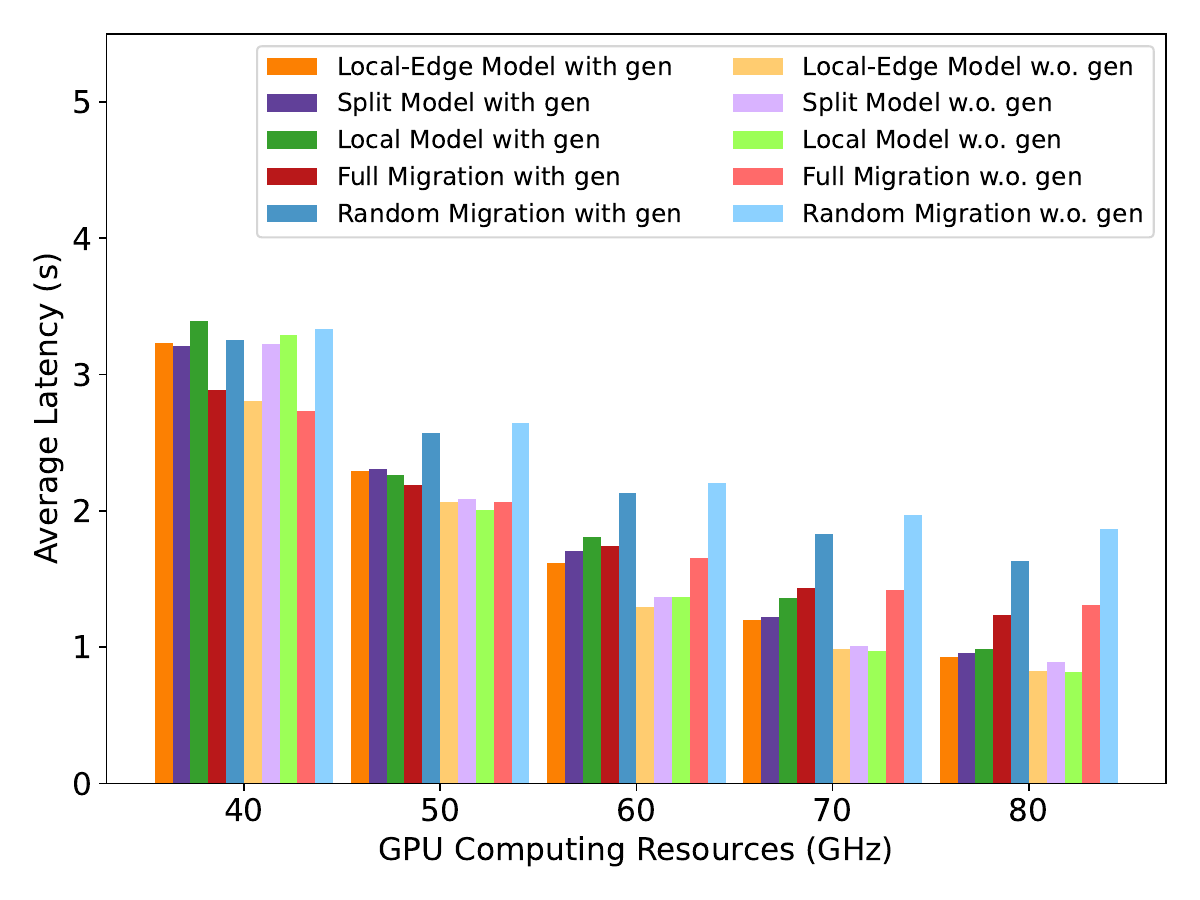}
    \caption{Total Latency versus different RSUs' GPU computing capabilities.}
    \label{fig:lat-bar}
\end{figure}

First, we investigate the effect of different RSU GPU computing resources on the average delay and assess the efficacy of our proposed scheme in minimizing average delay across different scenarios. Fig.~\ref{fig:lat-bar} shows that the proposed scheme consistently achieves lower average latency compared to other baseline schemes for RSU GPU computing resources ranging from 40 GHz to 80 GHz. This demonstrates the effectiveness of our approach in reducing average latency across various scenarios. The figure also shows that as the GPU computational resources of RSUs increase, the average delay correspondingly decreases. The reason is that more RSU GPU computation resources reduce the time required to increase RSU computation for VT tasks. Notably, as the RSU GPU computing resources increase, the advantage of \textit{Full Migration} keeps decreasing because when the GPU computing resources are sufficiently abundant, the pre-migrated RSUs can have enough resources to process the pre-migrated VT tasks, which further reduces the average latency. In addition, the patterns of the scheme with generated vehicles in the environment are similar to those of the scheme without generated vehicles in the environment, suggesting that the incorporated scheme with generated vehicles can effectively simulate the movement patterns of real vehicles, which in turn enables the algorithm to be trained in more diverse environments to improve the applicability of the algorithm.

\begin{figure}
    \centering
    \includegraphics[width=1\linewidth]{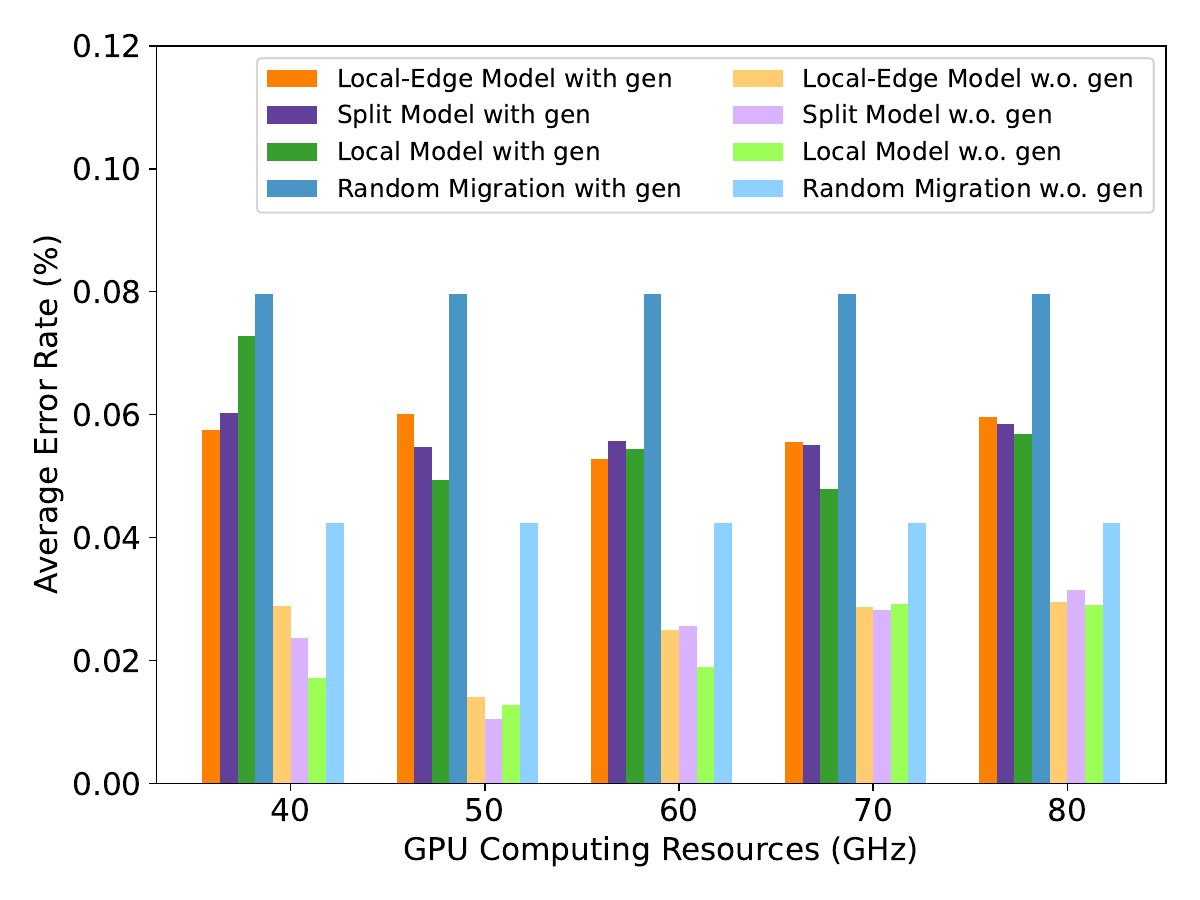}
    \caption{Average Error Rate versus different RSUs' GPU computing capabilities.}
    \label{fig:err-bar}
\end{figure}

\begin{figure}
    \centering
    \includegraphics[width=1\linewidth]{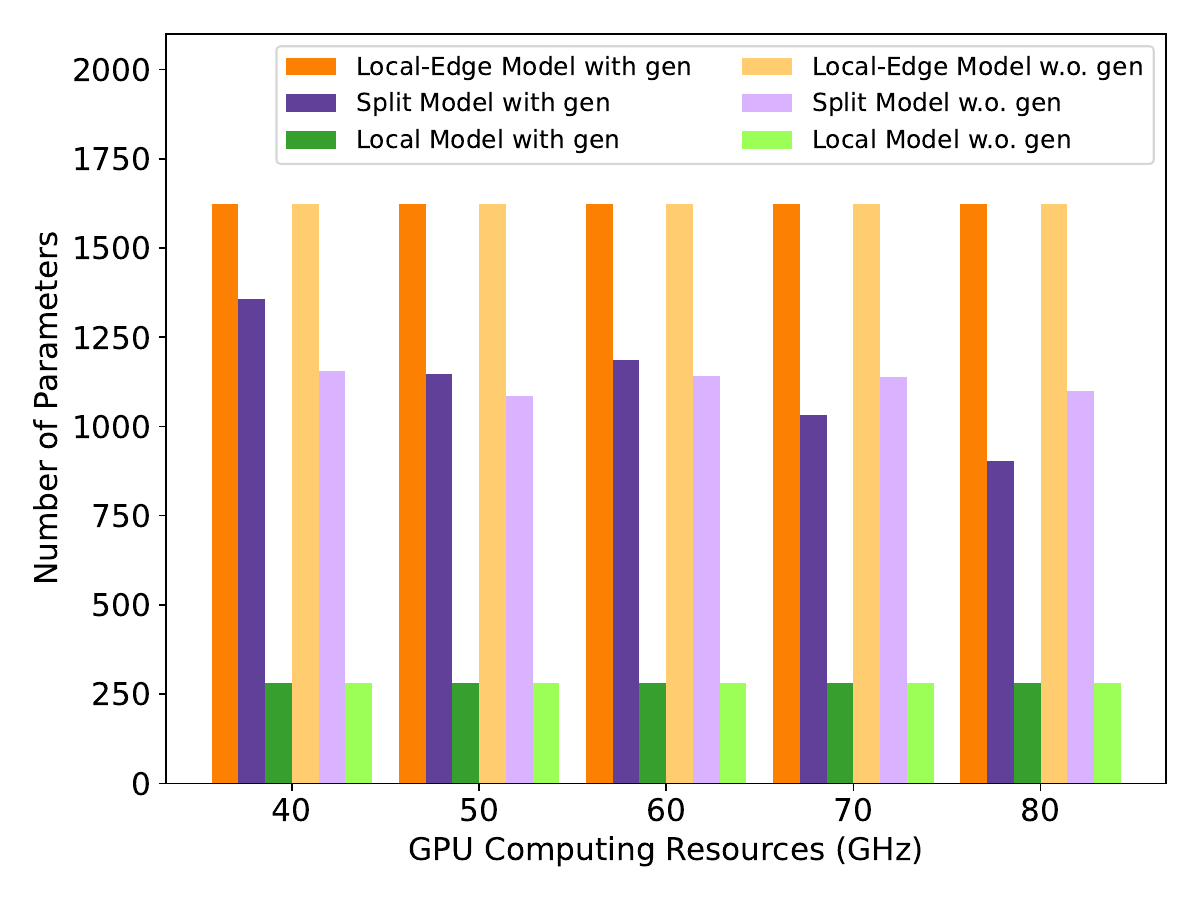}
    \caption{Computing Parameters versus different RSUs' GPU computing capabilities.}
    \label{fig:param-bar}
\end{figure}
Next, we evaluate the effectiveness of the proposed scheme to decrease the average error rate in different scenarios. As shown in Fig.~\ref{fig:err-bar}, for RSU GPU computational resources ranging from 40 GHz to 80 GHz, our proposed scheme consistently results in lower average error rates compared to the baseline scheme, which proves that the proposed scheme is effective in reducing the average error rates in different scenarios. In addition, the patterns of the schemes with generated vehicles in the environment are similar to the schemes without generated vehicles in the environment, which further indicates that the proposed scheme with the inclusion of generated vehicles can effectively simulate the movement patterns of real vehicles.

Finally, we evaluate the effectiveness of the proposed scheme in reducing the number of computational parameters in different scenarios. As shown in Fig.~\ref{fig:param-bar}, the proposed scheme reduces the number of computational parameters by 25\% compared to \textit{Local-Edge Model} for RSU GPU computational resources from 40 GHz to 80 GHz. Taking the previous results together, the proposed scheme maintains a similar performance to the \textit{Local-Edge Model} while reducing the number of computational parameters, striking a judicious balance between the number of computational parameters and the performance that outperforms the other baseline schemes. Therefore, the MSRL algorithm proves its superiority by not only converging to more strategic decisions, as indicated by higher average rewards but also achieving this goal with less computational resources. This dual benefit suggests our scheme could be widely applicable, particularly in VT migration environments where computational resources are limited and rapid decision-making is crucial.




\section{Conclusion}\label{conclusion}
In this paper, we have explored the optimization of VT migration to enhance immersive experiences within vehicular metaverses, addressing the challenges posed by vehicle mobility, dynamic RSU workloads, and resource diversity. By modeling the migration decision-making process as POMDP and introducing a novel VT migration decision framework, our scheme leverages split learning for efficient computational task distribution and MADRL for optimized migration decision-making. Furthermore, trajectory generation has been utilized to simulate realistic vehicle movements, thereby improving the algorithm's generalization capabilities across dynamic network environments. Finally, experimental results demonstrate that our scheme not only enhances the Quality of Experience by 29\% but also reduces the computational parameter count by approximately 25\% while maintaining similar performances, striking an effective balance between computational efficiency and immersive experience quality. 

Future work could consider partial migration techniques to selectively transfer only essential VT components, hierarchical edge-cloud architectures to reduce reliance on cloud-only storage, and predictive caching or adaptive data resolution strategies to handle varying vehicle speeds and large data sizes more efficiently. Such enhancements would further improve the scalability, responsiveness, and overall robustness of VT migration in real-world vehicular metaverse scenarios.



%

\ifCLASSOPTIONcaptionsoff
  \newpage
\fi



%
\bibliographystyle{IEEEtran}
\bibliography{main}

\begin{thebibliography}{10}
\providecommand{\url}[1]{#1}
\csname url@samestyle\endcsname
\providecommand{\newblock}{\relax}
\providecommand{\bibinfo}[2]{#2}
\providecommand{\BIBentrySTDinterwordspacing}{\spaceskip=0pt\relax}
\providecommand{\BIBentryALTinterwordstretchfactor}{4}
\providecommand{\BIBentryALTinterwordspacing}{\spaceskip=\fontdimen2\font plus
\BIBentryALTinterwordstretchfactor\fontdimen3\font minus \fontdimen4\font\relax}
\providecommand{\BIBforeignlanguage}[2]{{%
\expandafter\ifx\csname l@#1\endcsname\relax
\typeout{** WARNING: IEEEtran.bst: No hyphenation pattern has been}%
\typeout{** loaded for the language `#1'. Using the pattern for}%
\typeout{** the default language instead.}%
\else
\language=\csname l@#1\endcsname
\fi
#2}}
\providecommand{\BIBdecl}{\relax}
\BIBdecl

\bibitem{luo2023privacy}
X.~Luo, J.~Wen, J.~Kang, J.~Nie, Z.~Xiong, Y.~Zhang, Z.~Yang, and S.~Xie, ``Privacy attacks and defenses for digital twin migrations in vehicular metaverses,'' \emph{IEEE Network}, 2023.

\bibitem{li2023digital}
S.~Li, X.~Lin, J.~Wu, W.~Zhang, and J.~Li, ``Digital twin and artificial intelligence-empowered panoramic video streaming: Reducing transmission latency in the extended reality-assisted vehicular metaverse,'' \emph{IEEE Vehicular Technology Magazine}, 2023.

\bibitem{chen2024aiot}
J.~Chen, J.~Kang, M.~Xu, Z.~Xiong, D.~Niyato, C.~Chen, A.~Jamalipour, and S.~Xie, ``Multiagent deep reinforcement learning for dynamic avatar migration in aiot-enabled vehicular metaverses with trajectory prediction,'' \emph{IEEE Internet of Things Journal}, vol.~11, no.~1, pp. 70--83, 2024.

\bibitem{zhong2023blockchain}
Y.~Zhong, J.~Wen, J.~Zhang, J.~Kang, Y.~Jiang, Y.~Zhang, Y.~Cheng, and Y.~Tong, ``Blockchain-assisted twin migration for vehicular metaverses: A game theory approach,'' \emph{Transactions on Emerging Telecommunications Technologies}, vol.~34, no.~12, p. e4856, 2023.

\bibitem{hui2023digital}
Y.~Hui, Y.~Qiu, Z.~Su, Z.~Yin, T.~H. Luan, and K.~Aldubaikhy, ``Digital twins for intelligent space-air-ground integrated vehicular network: challenges and solutions,'' \emph{IEEE Internet of Things Magazine}, vol.~6, no.~3, pp. 70--76, 2023.

\bibitem{otoum2022feasibility}
S.~Otoum, N.~Guizani, and H.~Mouftah, ``On the feasibility of split learning, transfer learning and federated learning for preserving security in its systems,'' \emph{IEEE Transactions on Intelligent Transportation Systems}, 2022.

\bibitem{chen2021trajvae}
X.~Chen, J.~Xu, R.~Zhou, W.~Chen, J.~Fang, and C.~Liu, ``Trajvae: A variational autoencoder model for trajectory generation,'' \emph{Neurocomputing}, vol. 428, pp. 332--339, 2021.

\bibitem{luo2022estnet}
G.~Luo, H.~Zhang, Q.~Yuan, J.~Li, and F.-Y. Wang, ``Estnet: Embedded spatial-temporal network for modeling traffic flow dynamics,'' \emph{IEEE Transactions on Intelligent Transportation Systems}, vol.~23, no.~10, pp. 19\,201--19\,212, 2022.

\bibitem{wang2023trajectory}
X.~Wang, Z.~Jerome, C.~Zhang, S.~Shen, V.~V. Kumar, and H.~X. Liu, ``Trajectory data processing and mobility performance evaluation for urban traffic networks,'' \emph{Transportation Research Record}, vol. 2677, no.~3, pp. 355--370, 2023.

\bibitem{han2022dynamic}
Y.~Han, D.~Niyato, C.~Leung, D.~I. Kim, K.~Zhu, S.~Feng, X.~Shen, and C.~Miao, ``A dynamic hierarchical framework for iot-assisted digital twin synchronization in the metaverse,'' \emph{IEEE Internet of Things Journal}, vol.~10, no.~1, pp. 268--284, 2022.

\bibitem{xu2022full}
M.~Xu, W.~C. Ng, W.~Y.~B. Lim, J.~Kang, Z.~Xiong, D.~Niyato, Q.~Yang, X.~Shen, and C.~Miao, ``A full dive into realizing the edge-enabled metaverse: Visions, enabling technologies, and challenges,'' \emph{IEEE Communications Surveys \& Tutorials}, vol.~25, no.~1, pp. 656--700, 2022.

\bibitem{xu2023generative}
M.~Xu, D.~Niyato, J.~Chen, H.~Zhang, J.~Kang, Z.~Xiong, S.~Mao, and Z.~Han, ``Generative ai-empowered simulation for autonomous driving in vehicular mixed reality metaverses,'' \emph{IEEE Journal of Selected Topics in Signal Processing}, vol.~17, no.~5, pp. 1064--1079, 2023.

\bibitem{chen2023multiple}
J.~Chen, J.~Nie, M.~Xu, L.~Lyu, Z.~Xiong, J.~Kang, Y.~Tong, and W.~Jiang, ``Multiple-agent deep reinforcement learning for avatar migration in vehicular metaverses,'' in \emph{Companion Proceedings of the ACM Web Conference 2023}, 2023, pp. 1258--1265.

\bibitem{raja2022intelligent}
G.~Raja, N.~D. Philips, R.~K. Ramasamy, K.~Dev, and N.~Kumar, ``Intelligent drones trajectory generation for mapping weed infested regions over 6g networks,'' \emph{IEEE Transactions on Intelligent Transportation Systems}, 2022.

\bibitem{zheng2023differentiable}
H.~Zheng and R.~Mangharam, ``Differentiable trajectory generation for car-like robots with interpolating radial basis function networks,'' \emph{arXiv preprint arXiv:2303.00981}, 2023.

\bibitem{li2023trajectory}
J.~Li and W.~Zhao, ``Trajectory generation of ultra-low-frequency travel routes in large-scale complex road networks,'' \emph{Systems}, vol.~11, no.~2, p.~61, 2023.

\bibitem{wang2021large}
X.~Wang, X.~Liu, Z.~Lu, and H.~Yang, ``Large scale gps trajectory generation using map based on two stage gan,'' \emph{Journal of Data Science}, vol.~19, no.~1, pp. 126--141, 2021.

\bibitem{afrasiabi2022reinforcement}
S.~N. Afrasiabi, A.~Ebrahimzadeh, C.~Mouradian, S.~Malektaji, and R.~H. Glitho, ``Reinforcement learning-based optimization framework for application component migration in nfv cloud-fog environments,'' \emph{IEEE Transactions on Network and Service Management}, 2022.

\bibitem{dong2023dynamic}
P.~Dong, Z.~Duan, L.~Xia, and Z.~Li, ``A dynamic service migration method in mobile edge computing based on reinforcement learning,'' in \emph{2023 8th International Conference on Computer and Communication Systems (ICCCS)}.\hskip 1em plus 0.5em minus 0.4em\relax IEEE, 2023, pp. 456--462.

\bibitem{Lee1993Euclid}
J.~Lee, K.~Fukue, H.~Shimoda, and T.~Sakata, ``Grid data generation from contour images by using euclid distance transformation,'' in \emph{Proceedings of IGARSS '93 - IEEE International Geoscience and Remote Sensing Symposium}, 1993, pp. 1727--1729 vol.4.

\bibitem{shannon2001mathematical}
C.~E. Shannon, ``A mathematical theory of communication,'' \emph{ACM SIGMOBILE mobile computing and communications review}, vol.~5, no.~1, pp. 3--55, 2001.

\bibitem{ren2020edge}
P.~Ren, X.~Qiao, Y.~Huang, L.~Liu, C.~Pu, S.~Dustdar, and J.-L. Chen, ``Edge ar x5: An edge-assisted multi-user collaborative framework for mobile web augmented reality in 5g and beyond,'' \emph{IEEE Transactions on Cloud Computing}, vol.~10, no.~4, pp. 2521--2537, Oct. 2022.

\bibitem{hieu2023enabling}
N.~Q. Hieu, D.~T. Hoang, D.~N. Nguyen, and E.~Dutkiewicz, ``Enabling immersion and presence in the metaverse with over-the-air brain-computer interface,'' \emph{arXiv preprint arXiv:2303.10577}, 2023.

\bibitem{ji2016comparison}
Y.~Ji, H.~Liu, X.~Liu, Y.~Ding, and W.~Luo, ``A comparison of road-network-constrained trajectory compression methods,'' in \emph{2016 IEEE 22nd international conference on parallel and distributed systems (ICPADS)}.\hskip 1em plus 0.5em minus 0.4em\relax IEEE, 2016, pp. 256--263.

\bibitem{murphy2019map}
J.~Murphy, Y.~Pao, and A.~Yuen, ``Map matching when the map is wrong: Efficient on/off road vehicle tracking and map learning,'' in \emph{Proceedings of the 12th ACM SIGSPATIAL International Workshop on Computational Transportation Science}, 2019, pp. 1--10.

\bibitem{zhu2022knowledge}
Q.~Zhu, Y.~Chen, H.~Wang, Z.~Zeng, and H.~Liu, ``A knowledge-enhanced framework for imitative transportation trajectory generation,'' in \emph{2022 IEEE International Conference on Data Mining (ICDM)}.\hskip 1em plus 0.5em minus 0.4em\relax IEEE, 2022, pp. 823--832.

\bibitem{DelinGuo2020MAPPO}
D.~Guo, L.~Tang, X.~Zhang, and Y.-C. Liang, ``Joint optimization of handover control and power allocation based on multi-agent deep reinforcement learning,'' \emph{IEEE Transactions on Vehicular Technology}, 2020.

\bibitem{yuan2010t}
J.~Yuan, Y.~Zheng, C.~Zhang, W.~Xie, X.~Xie, G.~Sun, and Y.~Huang, ``T-drive: driving directions based on taxi trajectories,'' in \emph{Proceedings of the 18th SIGSPATIAL International conference on advances in geographic information systems}, 2010, pp. 99--108.

\end{thebibliography}

%




\end{document}